\documentclass[prl]{revtex4}

\usepackage[pdftex]{graphicx}
\usepackage{amssymb,amsfonts,amsmath}
\usepackage{graphicx,amsmath}

\begin{document}

\title{An incompressible state of a photo-excited electron gas}
\author{Alexei D. Chepelianskii$^{(a,b)}$, Masamitsu Watanabe$^{(c)}$, Kostyantyn Nasyedkin$^{(d)}$, Kimitoshi Kono$^{(d,e,f)}$ and Denis Konstantinov$^{(g)}$ \\
$(a)$ LPS, Universit\'e Paris-Sud, CNRS, UMR 8502, F-91405, Orsay, France (email alexei.chepelianskii@u-psud.fr) \\
$(b)$ Cavendish Laboratory, University of Cambridge, J J Thomson Avenue, Cambridge CB3 OHE, UK\\
$(c)$ Low Temperature Physics Laboratory, RIKEN, Wako, Saitama 351-0198, Japan\\
$(d)$ Quantum Condensed Phases Research Team, RIKEN CEMS, Wako, Saitama 351-0198, Japan\\
$(e)$ Institute of Physics, National Chiao Tung University, Hsinchu 30010, Taiwan\\
$(f)$ Institute of Physics, Kazan Federal University, Kazan 420008, Russia\\
$(g)$ Okinawa Institute of Science and Technology, Onna, Okinawa 904-0412, Japan
}

\pacs{Physical sciences/Physics/Condensed-matter physics/Quantum fluids and solids and Quantum Hall} 

\begin{abstract}
Two dimensional electrons in a magnetic field can form new states of matter characterized by
topological properties and strong electronic correlations as displayed in the integer and fractional quantum Hall states.
In these states the electron liquid displays several spectacular characteristics which manifest themselves in transport experiments 
with the quantization of the Hall resistance and a vanishing longitudinal conductivity
or in thermodynamic equilibrium when the electron fluid becomes incompressible. 
Several experiments have reported that dissipation-less transport can be 
achieved even at weak, non-quantizing magnetic fields when the electrons absorb photons 
at specific energies related to their cyclotron frequency.
Compressibility measurements on electrons on liquid helium demonstrate 
the formation of an incompressible electronic state under these resonant excitation conditions.
This new state provides a striking example of irradiation induced self-organization in a quantum system.
\end{abstract}

\maketitle

{\it Introduction}

The discovery of the integer and fractional quantum Hall effects  \cite{Klitzing,Tsui,Jain,Novoselov,Kim2005,Kim2009} revealed 
the existence of new states of matter characterized by topological properties and strong electronic correlations triggering an intense theoretical 
and experimental research activity. These efforts lead to a detailed microscopic understanding of 
the main experimental phenomena and to some of the most beautiful conceptual breakthroughs in condensed matter physics \cite{Ezawa}.
The observation of a new dissipationless transport regime at low magnetic fields under microwave irradiation \cite{mani2002,zudov2003} raised a new challenge regarding our understanding of two dimensional electron systems. Microwave induced zero resistance states (ZRS) 
appear at high microwave excitation powers when the ratio between the photon energy $\hbar \omega$ and Landau level spacing $\hbar \omega_{{\rm c}}$ 
is close to a value within the fraction series: $J=\omega/\omega_{{\rm c}} = 1+1/4,2+1/4, ...$. 
In theory however, the motion of electrons in a magnetic field in ultra-clean samples 
is well described as a harmonic oscillator where the selection rules only allow transitions between nearby oscillator states corresponding to $J = 1$. 
The theoretical explanations proposed so far, have attempted to resolve this contradiction by considering the role of sharp inhomogeneities
due to a short-range disorder potential  \cite{Girvin,Dmitriev,review}, edges \cite{ToulouseEdge,Levin} and contacts \cite{Mikhailov}. 
The dominant microscopic picture for ZRS is currently an ensemble of domains with vanishing local conductivity \cite{review},
but the formation of a collective state with long range order has also been suggested  \cite{ToulouseS1,ToulouseS2}.
So far, the experimental evidence does not provide a definite proof in support of one of the available models
and despite intense experimental efforts \cite{West2004,Smet,Zudov2006,Mani2009,Wiedmann,Dorozhkin,Mani2011} the microscopic nature of ZRS remains a puzzle. One of the difficulties is that while the manifestations of ZRS in transport phenomena are spectacular,
such as vanishing longitudinal resistance, other indications of these novel electronic states remain elusive.

The observation of ZRS for surface electrons on helium under intersubband excitation \cite{Denis2,Denis1} has opened a new research direction in this field since the strong Coulomb interactions in this system allow collective effects, such as for example Wigner crystallization, to be observed more readily \cite{andrei,monarkha_book}. Previously, we reported a strong redistribution of the electron density under irradiation which coincided with the appearance of ZRS \cite{DenisTlse}, although the underlying mechanism was not elucidated. Since the formation of ZRS in a Hall system coincides with vanishing conductivity, the observed redistribution may simply be a consequence of the expected long charge relaxation rates in this regime.
In the experiments presented here, we systematically study the behaviour of the electronic density under irradiation and demonstrate a regime in which electrons stabilize at a fixed steady-state density independent of their initial density profile and the electrostatic confinement potential.
Since in this regime the electron density is not changed by an increase of the holding electrostatic forces which tend to compress the electron cloud, we describe this new phase of the electron gas as an incompressible state.

{\it Results}

\begin{figure}
  \centerline{\includegraphics[clip=true,width=7cm]{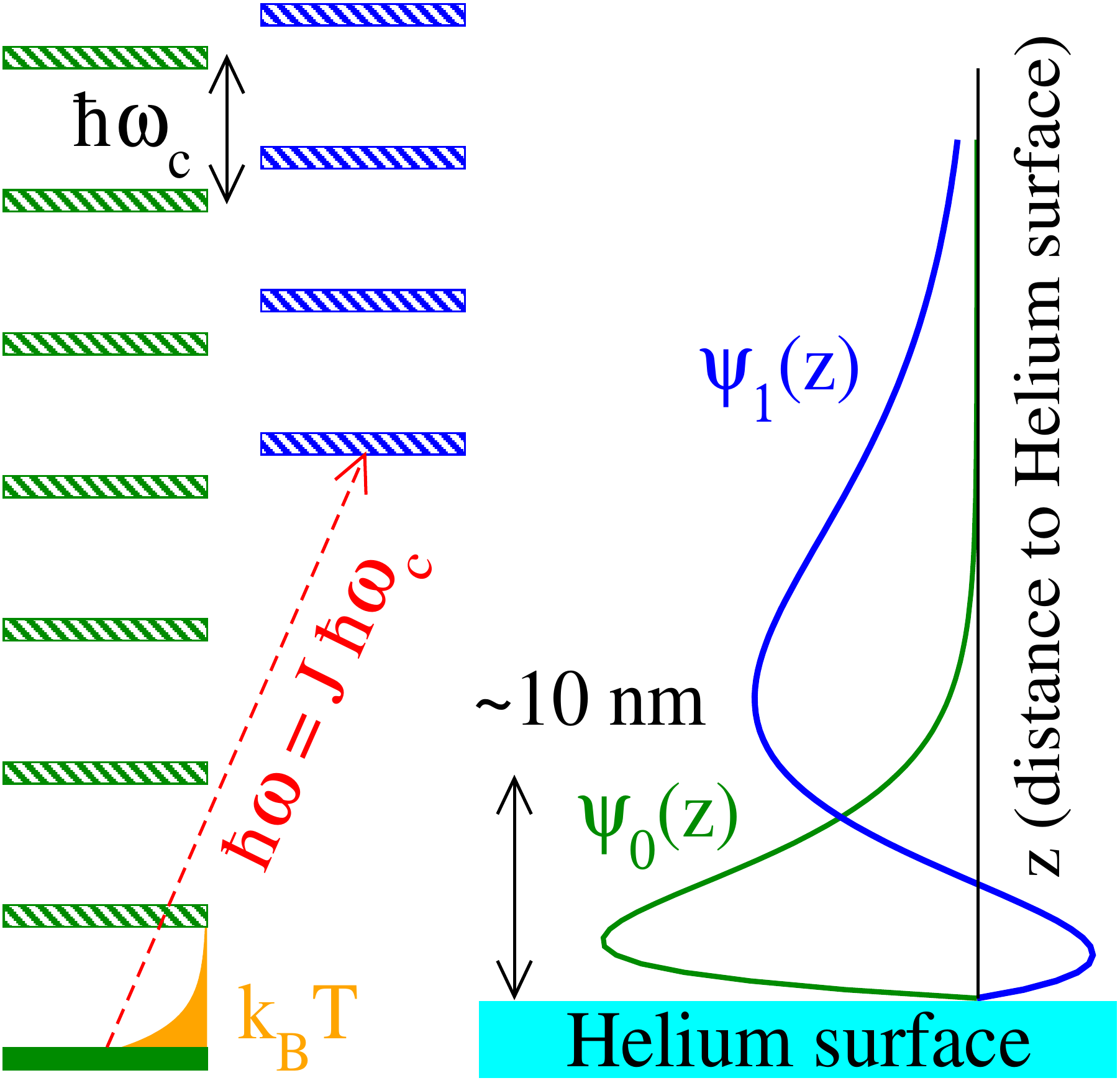}}
\caption{{\bf Accessible energy levels.} Level diagram of the accessible states for the surface electrons in our experiment.
Several bound states are formed owing to the attractive image charge potential created by the helium surface,
the ground state and the first excited state are separated by energy $\hbar \omega$. Free motion along the helium surface transforms these states 
into conduction bands which under a quantizing magnetic field split into a manifold of discrete Landau levels. 
The ground state manifold is characterized by a wavefunction  $\psi_0(z)$ that is  
localized close to the helium surface whereas the wavefunction of the excited state manifold is centered farther away from the helium surface. 
}
\label{FigLevels}
\end{figure}

\begin{figure}
  \centerline{\includegraphics[clip=true,width=12cm]{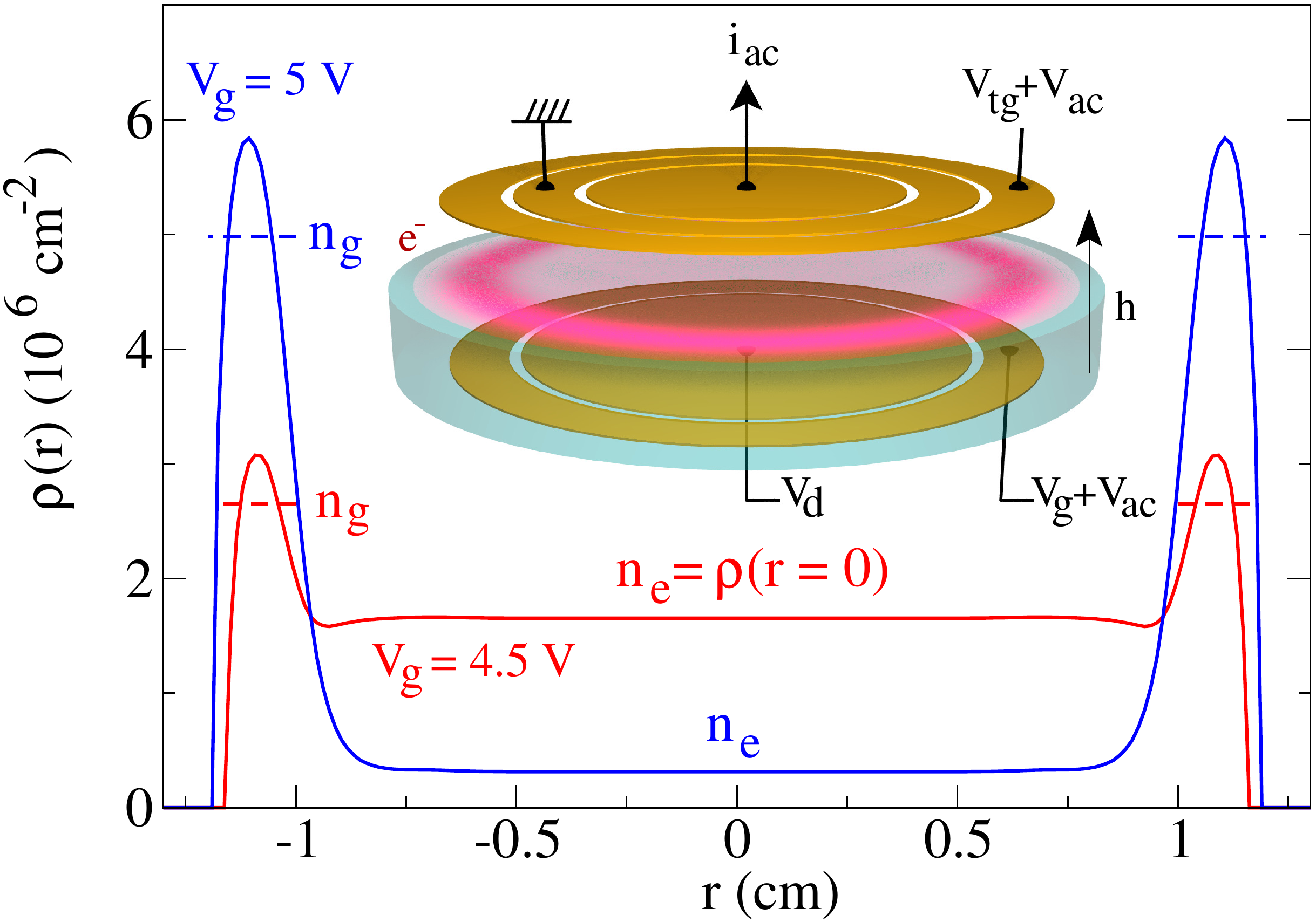}}
\caption{{\bf Electron distribution inside the cell.} Theoretical electron density profiles $\rho(r)$ for different values of the guard potential $V_{{\rm g}}$ for a cloud with 
$N_{{\rm e}} = 8\;\times 10^6$ trapped electrons and corresponding values of the mean electron density in the center $n_{{\rm e}}$ 
and in the guard $n_{{\rm g}}$. The inset illustrates the helium cell geometry in our experiment 
and the electrostatic potentials applied to the electrodes during the compressibility experiments. 
The top of the cell is split in three circular electrodes with radii $R_{{\rm i}} = 0.7, R_{{\rm d}} = 1, R_{{\rm g}} = 1.3\;{\rm cm}$, and the height of the cell is $h = 2.6\;{\rm mm}$.
For compressibility measurements, an AC driving potential is applied to both top and bottom guard electrodes and the induced image charge current $i_{{\rm ac}}$ is measured on the central top electrode. 
}
\label{FigSetup}
\end{figure}

{\bf Description of the system}: A population of electrons us trapped on a liquid helium surface forming a nondegenerate two-dimensional electron gas. 
The energy levels accessible to the surface electrons in a quantizing perpendicular magnetic field are shown in Fig.~\ref{FigLevels}.
They are formed by Landau levels separated by energy $\hbar \omega_{{\rm c}}$ and intersubband excitations of energy $\hbar \omega$ perpendicular 
to the helium surface \cite{andrei,monarkha_book}. Our experiments are performed at a temperature of $T = 300\;{\rm mK}$ 
much smaller than the Landau level spacing $k_{{\rm B}} T \ll \hbar \omega_{{\rm c}}$ so that in equilibrium the electrons mainly fill the lowest Landau level, whereas under resonant 
irradiation at energy $\hbar \omega$, they can be excited into another subband manifold \cite{Denis2,Denis1}. 
Note that the two level system formed by the two subbands has been proposed as a candidate system for quantum computing \cite{Dykman1999}.
Spatially, the electrons are distributed between two regions on the helium surface (see Fig.~\ref{FigSetup}), a central region above the 
disc-shaped electrode at potential $V_{{\rm d}}$ and a surrounding guard region above the ring electrode 
held at potential $V_{{\rm g}}$. We denote $n_{{\rm e}}$ and $n_{{\rm g}}$ as the mean electron densities in the central and guard regions respectively. 
As in a field effect transistor $n_{{\rm e}}$ and $n_{{\rm g}}$ can be controlled by changing the potentials $V_{{\rm d}}$ and $V_{{\rm g}}$. The
key difference here is that for surface electrons the total number of electrons $N_{{\rm e}}$ in the cloud is fixed as long
as the positive potentials $V_{{\rm d}}$ and $V_{{\rm g}}$ are sufficiently strong to balance the electron-electron Coulomb repulsion.
Examples of simulated electron density profiles for different values of $V_{{\rm g}}$ are shown in Fig.~\ref{FigSetup},
the simulations were performed within an electrostatic model as described in \cite{Wilen,Kovdrya,Fabien2014}.

In addition to controlling the density profile of the electron cloud $\rho(r)$, the potentials $V_{{\rm d}}$ and $V_{{\rm g}}$ also change the perpendicular holding field in the cell $E_z$. To avoid this unwanted effect in our experiments,
we fixed the value of $V_{{\rm d}}$ and changed the potential $V_{{\rm tg}}$ simultaneously with $V_{{\rm g}}$ keeping the difference $V_{{\rm tg}} - V_{{\rm g}}$ 
between top and bottom guard electrode voltages equal to $V_{{\rm d}}$. This choice ensures a uniform value of $E_z$ across the cell. Since lowering the potential $V_{{\rm g}}$ compresses the electron cloud towards the center of the helium cell, we define the compressibility of the electron system as $\chi = -d n_{{\rm e}}/d V_{{\rm g}}$. In this definition, the electrostatic potential plays the role of the chemical potential in quantum Hall systems. This difference is due to the non-degenerate statistics for electrons on helium for which the Fermi energy is much smaller than the thermal energy.

{\bf Compressibility in equilibrium}: we developed the following method to measure the compressibility of the electron cloud.
An AC voltage excitation with amplitude $V_{{\rm ac}} = 25\;{\rm mV}$ was applied on the top and bottom guard electrodes (see Fig.~\ref{FigSetup}) at a low frequency $f_{{\rm ac}} \simeq 2\;{\rm Hz}$, for which the electron density quasi-statically follows the driving potential.
The induced modulation of $n_{{\rm e}}$  was measured by recording the AC current $i_{{\rm ac}}$
created by the motion of image charges on the top central electrode with radius $R_{{\rm i}} = 0.7\;{\rm cm}$.
The correspondence between the variation of $n_{{\rm e}}$ and $i_{{\rm ac}}$ was established using plane capacitance electrostatics, for which an electron trapped 
at the middle of the cell induces half an image charge of $e/2$ on the top electrode.
The use of a simplified electrostatic model is justified here, since the gradients of $\rho(r)$ are located 
away from the electrode on which $i_{{\rm ac}}$ is measured. This leads to the following expression for the compressibility $\chi =  -d n_{{\rm e}}/d V_{{\rm g}}$:
\begin{align}
\chi = \frac{i_{{\rm ac}}}{e \pi^2 f_{{\rm ac}} R_{{\rm i}}^2 V_{{\rm ac}}} 
\label{chiac}
\end{align}
The presented measurement technique has a strong similarity to that used in compressibility experiments in the quantum Hall regime \cite{Tessmer}, providing an additional justification for our definition of compressibility. In supplementary note 1, we provide a more thorough discussion on our definition of compressibility and give a detailed derivation of Eq.~(\ref{chiac}) which is explained in supplementary note 2.

\begin{figure}
\centerline{\includegraphics[clip=true,width=12cm]{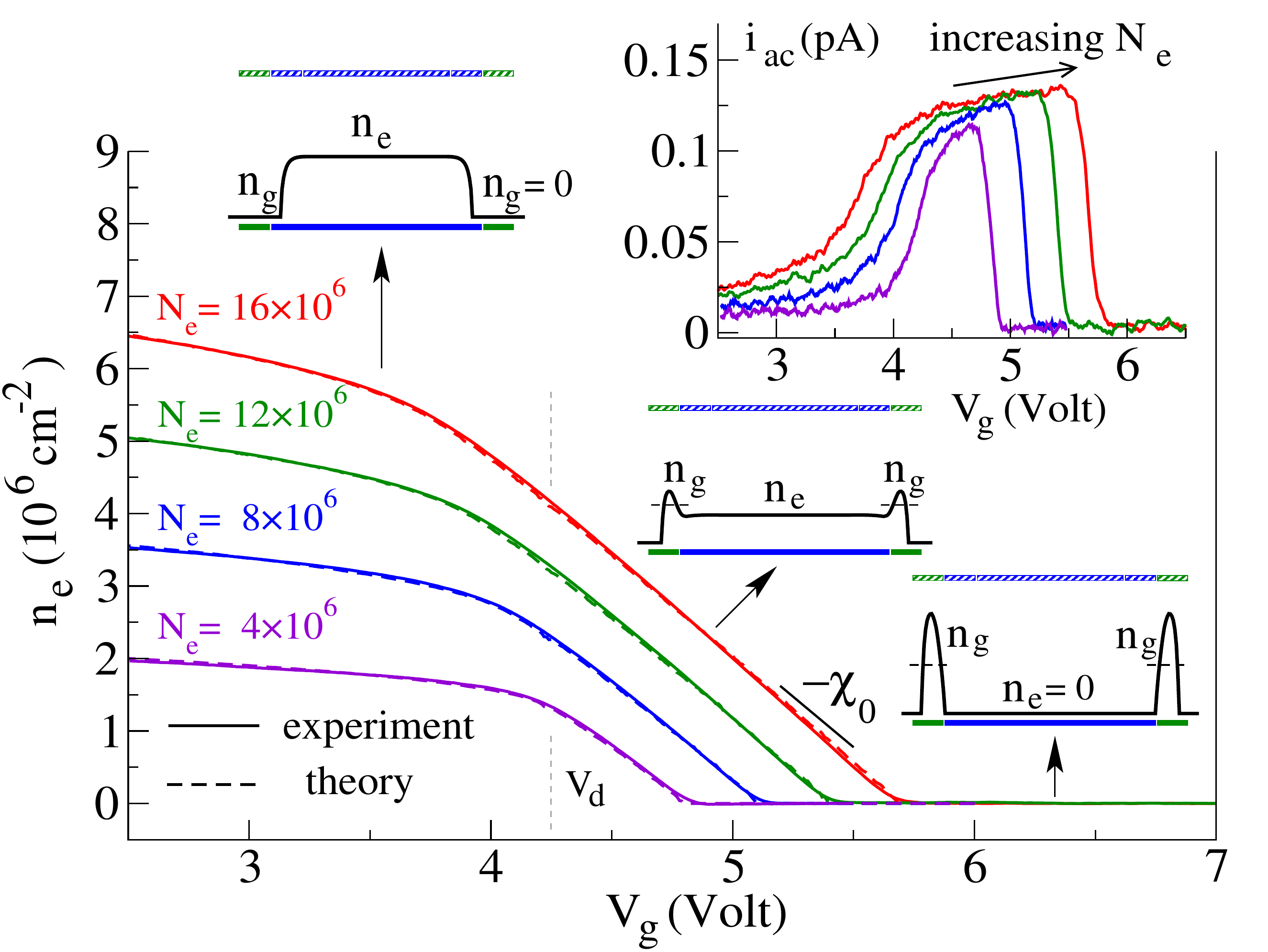}} 
\caption{{\bf Density without irradiation.} Dependence of $n_{{\rm e}}$ on the guard potential $V_{{\rm g}}$ without irradiation obtained using Eq.~(\ref{chiac}) for 
different numbers of trapped electrons $N_{{\rm e}}$, the experimental values for the current are shown in the inset. 
The total number of electrons in the cloud is determined by comparing the measured dependence $n_{{\rm e}}(V_{{\rm g}})$ 
with theoretical calculations based on electrostatic modeling of the electron cloud.
The sketches illustrate the typical electron density profiles along the $n_{{\rm e}}(V_{{\rm g}})$ curves. 
A very good agreement between experiment and modeling is achieved using the total electron number 
$N_{{\rm e}}$ as the only fitting parameter. Over a large range of $V_{{\rm g}}$ and $N_{{\rm e}}$ the compressibility is well 
approximated by the constant value $\chi_0 \simeq 2.9 \times 10^6 \;{\rm cm^{-2} V^{-1}}$.
Temperature was $T = 300\;{\rm mK}$, we verified
that as expected these measurements were independent of the magnetic field.
}
\label{FigDensity}
\end{figure}

Using Eq.~(\ref{chiac}), the dependence $n_{{\rm e}}(V_{{\rm g}})$ 
can be reconstructed by integrating $\chi$ with respect to $V_{{\rm g}}$ starting from the high $V_{{\rm g}}$ limit where $n_{{\rm e}} = 0$. 
The obtained results are illustrated in Fig.~\ref{FigDensity} which shows $i_{{\rm ac}}$ and the corresponding density 
$n_{{\rm e}}$ as a function of $V_{{\rm g}}$ for several $N_{{\rm e}}$ values. The experimental curves for $n_{{\rm e}}(V_{{\rm g}})$ are 
compared with the results of electrostatic simulations \cite{Wilen,Kovdrya,Fabien2014,Badrutdinov} that use the total electron number $N_{{\rm e}}$ as the single fitting parameter for each of the obtained curves. 
The simulations exhibit an extremely good agreement with the experimental results. 
The dependence $n_{{\rm e}}(V_{{\rm g}})$ can be understood as follows: a large positive potential $V_{{\rm g}}$ attracts the electrons towards the guard electrodes, whereas at low $V_{{\rm g}}$ the electrons are repelled from the guard region and concentrate at the center of the cell. 
The potential from the guard electrodes is then almost completely screened, leading to a value of $n_{{\rm e}}$ that is almost independent of $V_{{\rm g}}$. 
At intermediate $V_{{\rm g}}$, electrons occupy both the central and guard regions. In this case $n_{{\rm e}}$ decreases linearly with increasing $V_{{\rm g}}$ until the central region becomes completely depleted. At this point $n_{{\rm e}} = 0$ regardless of the value of $V_{{\rm g}}$. 
In the intermediate regime the compressibility $\chi = -\frac{d n_{{\rm e}}}{d V_{{\rm g}}}$ depends only weakly on the total number of trapped electrons $N_{{\rm e}}$ 
and the values of the confining potentials. This observation can be understood by considering a simplified electrostatic model in which the two electron reservoirs in the disc and guard 
are treated as plane capacitors. This model, which is presented in more detail in supplementary note 3, leads to a value for $\chi$ that depends only 
on the geometrical cell parameters (provided the disc/guard reservoirs are not empty): 
\begin{align}
\chi_0 = -\frac{d n_{{\rm e}}}{d V_{{\rm g}}} = \frac{4 \epsilon_0}{e h(1 + S_{{\rm d}}/S_{{\rm g}})} 
\end{align}
here, $h = 2.6\;{\rm mm}$ is the cell height, $S_{{\rm d}} = \pi R_{{\rm d}}^2$ and $S_{{\rm g}} = \pi (R_{{\rm g}}^2 - R_{{\rm d}}^2)$ are the surface areas of the bottom disc/guard electrodes 
($R_{{\rm g}}$ being the outer radius of the guard electrodes) and $\epsilon_0$ is the vacuum permittivity.  
We experimentally find $\chi_0 \simeq 2.9 \times 10^{6}\;{\rm cm^{-2} V^{-1}}$ in good agreement with the estimation obtained 
using the geometrical cell parameters $R_{{\rm d}} = 1\;{\rm cm}$ and $R_{{\rm g}}  = 1.3\;{\rm cm}$. This reference value will be used 
to normalize the compressibility in our following experiments.

{\bf Compressibility under irradiation}: We next present our compressibility measurements in the presence of microwave irradiation, focusing on the ZRS fraction 
$J = \omega/\omega_{{\rm c}} = 6.25$, with $\omega = 2 \pi \times 139\;{\rm GHz}$ and a magnetic field of $B = 0.79\;{\rm Tesla}$. Outside ZRS regions the compressibility is not changed by microwave irradiation since it is independent of the conductivity of the system $\sigma_{{\rm xx}}$ provided it remains finite. The holding field, identical in the central and guard regions ($V_{{\rm d}} = V_{{\rm g}} - V_{{\rm tg}} = 4.24\;{\rm V}$), was chosen to tune the photon energy in resonance with the intersubband transition.

\begin{figure}
\centerline{\includegraphics[clip=true,width=15cm]{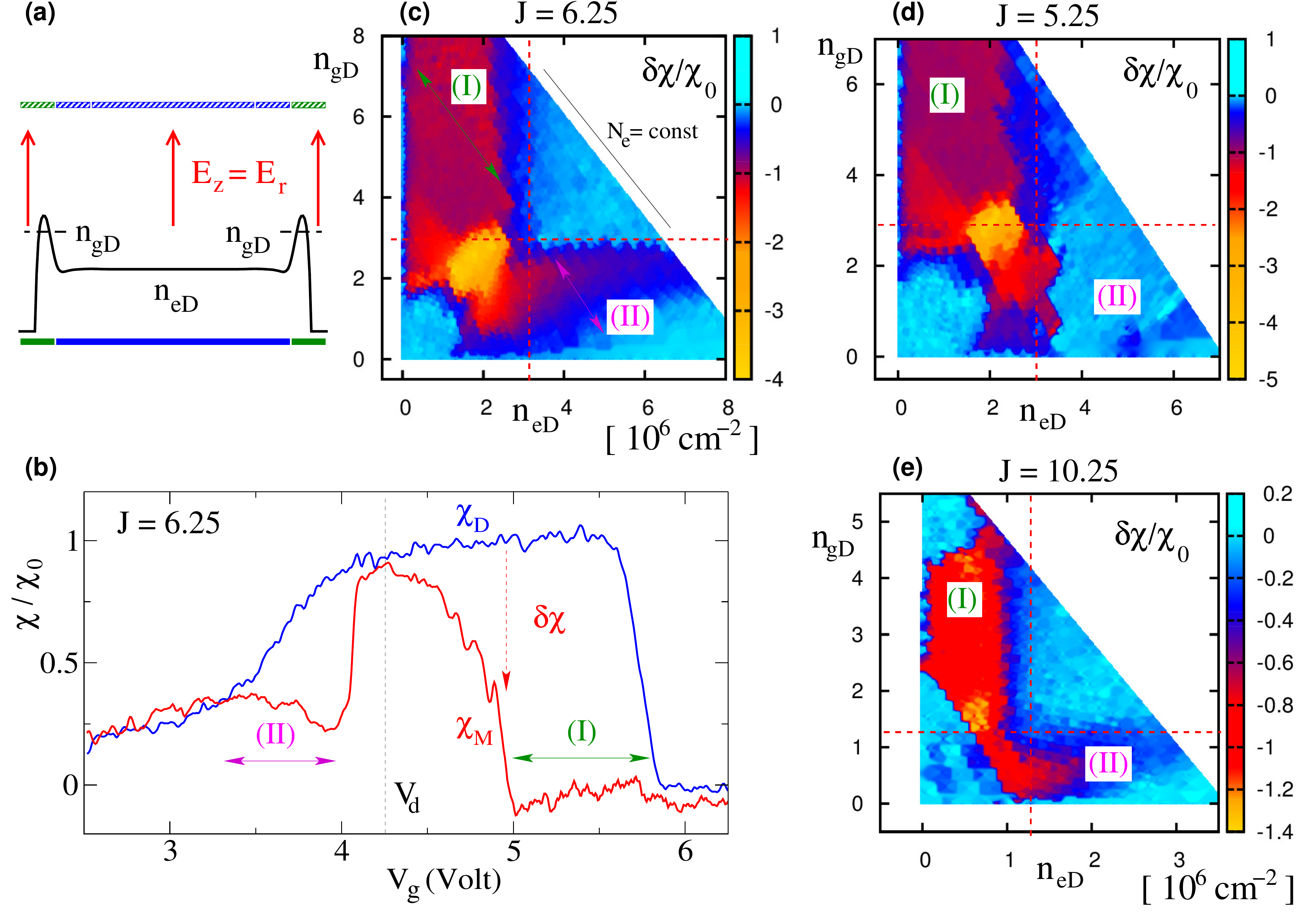}}
\caption{{\bf Compressibility under irradiation.} Compressibility $\chi$ as a function of guard voltage $V_{{\rm g}}$ in the dark ($\chi_{{\rm d}}$) and under microwave irradiation ($\chi_M$) at 
a frequency of $\omega = 2\pi\times 139\;{\rm GHz}$ and a magnetic field $B = 0.79\;{\rm Tesla}$ corresponding to $J = \omega/\omega_{{\rm c}} = 6.25$. 
As shown in ($\mathbf{a}$), the perpendicular electric field was fixed to the intersubband resonance value $E_r$ in both the central and guard regions. 
Panel ($\mathbf{b}$) shows the compressibility under microwave irradiation  for $N_{{\rm e}} = 18\times10^6$, $\chi$ changes under irradiation 
in two distinct regions: region $(I)$ where $\chi_M$ almost vanishes under irradiation and region $(II)$ where $\chi_M$ is significantly reduced.
($\mathbf{c}$) displays $\delta \chi/\chi_0$ measured at different $N_{{\rm e}}$ values as a function of the equilibrium density at the center $n_{{\rm eD}}$ and in the guard $n_{{\rm gD}}$. The boundaries of the anomalous regions $(I)$ and $(II)$ 
are aligned along lines of constant density $n_{{\rm eD}}$ and $n_{{\rm gD}}$ respectively. 
The color scale indicates the change in the compressibility under irradiation, the value $\delta \chi/\chi_0 = -1$ (red color)
corresponding to incompressible states. The positions of the colored arrows corresponds to the trace in panel $(\mathbf{b})$.
($\mathbf{d},\mathbf{e}$) show results similar to panel $(\mathbf{c})$ but obtained in two other ZRS regions: $J = 5.25$ and $J = 10.25$. The vertical/horizontal boundaries coincide approximately at $J = 5.25$ and $J = 6.25$ but are located at significantly lower densities for $J = 10.25$.
}
\label{FigChiMW}
\end{figure}

The compressibilities measured in the dark $\chi_{{\rm d}}$ and under irradiation $\chi_M$ are compared in Fig.~\ref{FigChiMW}.
Two singular regions where  $\chi_M$ and $\chi_{{\rm d}}$ differ are present: 
region $(I)$ at high $V_{{\rm g}}$ and region $(II)$ at low $V_{{\rm g}}$.
Hereafter on we use different notations for the average densities in the dark $n_{{\rm eD}}$ and $n_{{\rm gD}}$ and under microwave irradiation $n_{{\rm eM}}$ and $n_{{\rm gM}}$ to avoid ambiguity.
In region $(I)$, the dark density in the guard is higher than at the center $n_{{\rm gD}} > n_{{\rm eD}}$, whereas in region $(II)$ we have $n_{{\rm gD}} < n_{{\rm eD}}$.
In both regions, a strong suppression of the compressibility is observed and $\chi_M$ strikingly vanishes in most of region $(I)$.
In contrast, at $V_{{\rm g}} = V_{{\rm d}} = 4.24\;{\rm V}$ where the electrons are distributed evenly between center and the guard,
the compressibility remains almost unchanged under irradiation: $\chi_M \simeq \chi_{{\rm d}} \simeq \chi_0$.

To clarify the physical origin of the anomalous regions $(I)$ and $(II)$, we convert the experimentally 
controlled variables $V_{{\rm g}}$ and $N_{{\rm e}}$ to physically more relevant densities $n_{{\rm eD}}$ and $n_{{\rm gD}}$. 
We obtained $n_{{\rm eD}}$ and $N_{{\rm e}}$ from the compressibility measurements in the dark (as in Fig.~\ref{FigDensity}). 
The quantity $n_{{\rm gD}}$ could not be measured directly in a reliable manner owing to the unavoidable effect
of the density gradients in the guard region. We thus calculated $n_{{\rm gD}}$ by averaging the simulated density profiles $\rho(r)$ over the guard region.
This procedure is justified by the excellent agreement between the compressibility measurements 
and our numerical simulations that we demonstrated in absence of irradiation.

Using this method we summarize on the $n_{{\rm eD}}, n_{{\rm gD}}$ plane the changes in compressibility $\delta \chi = \chi_M - \chi_{{\rm d}}$ under irradiation 
measured at different values of $V_{{\rm g}}$ and $N_{{\rm e}}$ while fixing all the other parameters (magnetic field, microwave frequency and power, temperature and perpendicular electric field; the dependence on microwave power and magnetic field is shown in supplementary figures 1-5). The results are shown in the color-scale panels in Fig.~\ref{FigChiMW}. The anomalous regions $(I)$ and $(II)$ are upper bounded by lines of constant density in the center $n_{{\rm eD}} = n_{{\rm c}}$ and in the guard $n_{{\rm gD}} = n_{{\rm c}}$ where we introduced $n_{{\rm c}} \simeq 3 \times 10^6\;{\rm cm^{-2}}$.
Indeed for $n_{{\rm gD}} > n_{{\rm c}}$ and $n_{{\rm eD}} > n_{{\rm c}}$ the change in compressibility $|\delta \chi| \ll \chi_0$ is negligible and the electron
density is still described by the electrostatics of the gas phase. Similarly at low densities: $n_{{\rm eD}}$ and $n_{{\rm gD}} < 1.5 \times 10^6 \;{\rm cm^{-2}}$,
we also find no deviations from the equilibrium $\chi$ values. The incompressible regions $(I)$ and $(II)$ occupy the space $n_{{\rm eD}} < n_{{\rm c}} < n_{{\rm gD}}$ 
and a fraction of the space $n_{{\rm gD}} < n_{{\rm c}} < n_{{\rm eD}}$, they are characterized by $\delta \chi / \chi_0 \simeq -1$ on Fig.~(\ref{FigChiMW}). 
Finally, in the remaining area on the $n_{{\rm eD}}, n_{{\rm gD}}$ plane, $\chi$ becomes negative under irradiation for example at $n_{{\rm eD}} = n_{{\rm gD}} = 2 \times 10^6 \;{\rm cm^{-2}}$. 
To highlight the robustness of our results, Fig.~\ref{FigChiMW} also shows similar data obtained at two other ZRS fractions: $J = 5.25$ and $J = 10.25$. Incompressible regions appear for these cases as well, we note that for $J = 10.25$ the position of the boundary $n_{{\rm c}}$ is displaced towards significantly lower densities $n_{{\rm c}} \simeq 1.3 \times 10^{6} \;{\rm cm^{-2}}$.

The incompressible regions with vanishing $\chi$ correspond to an unexpected regime where the 
density becomes independent of the compressing confinement potential $V_{{\rm g}}$. 
In the integer quantum Hall effect, incompressible phases appear owing to the finite energy required to add electrons to 
a system in which the Landau levels available at the Fermi energy are all fully occupied.
This explanation, however, is not applicable to electrons on helium since they form a non-degenerate electron gas.
Experiments on the quantum Hall effect have also shown that a vanishing longitudinal conductivity $\sigma_{{\rm xx}}$,
can freeze a non-equilibrium electron density distribution since the charge relaxation time-scales can become exponentially large \cite{dolgoporov,jeannert}.
An explanation based on the vanishing conductivity $\sigma_{{\rm xx}}$  seems natural owning to the coincidence between the onset of charge redistribution and ZRS. We can determine experimentally whether the incompressible behavior can be explained only on the basis of $\sigma_{{\rm xx}} = 0$.
Indeed a state with $\sigma_{{\rm xx}} = 0$ is expected to freeze the existing density distribution due to the very long charge relaxation rates;
thus, the final state should depend in a non trivial way on the equilibrium density profile and on the kinetics of the transition to ZRS. 

\begin{figure}
\centerline{\includegraphics[clip=true,width=15cm]{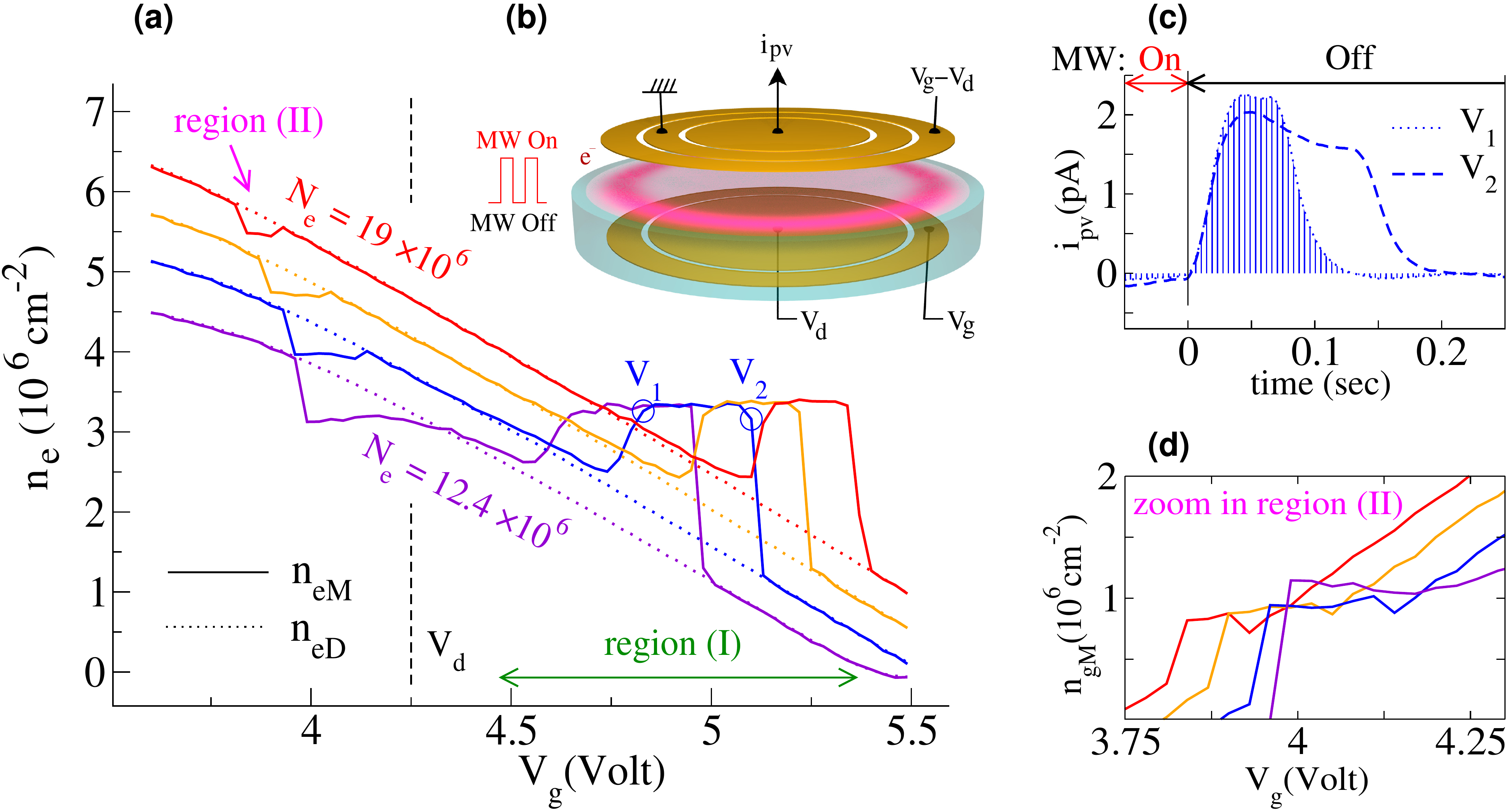}}
\caption{{\bf Photo-current density measurements.}  ({\bf a}) Density under irradiation as determined from transient photo-current measurements.
The electron cloud was prepared at different initial density distributions by changing $V_{{\rm g}}$ and performing experiments at several values of $N_{{\rm e}} = 12.4, 14.5, 16.8, 19 \; [ \times 10^{6} ] $ . The dependence of the equilibrium density $n_{{\rm eD}}$ on $V_{{\rm g}}$ was determined using the method from Fig.~2, 
it is indicated by  dotted lines. For each initial condition, we then applied a sequence of on/off microwave 
pulses and recorded the photocurrent $i_{{\rm pv}}$ induced by the electron redistribution. The measurement geometry is shown in ({\bf b}). Two $i_{{\rm pv}}(t)$ traces are shown in ({\bf c}) for two $V_{{\rm g}}$ values noted $V_1 = 4.83\;{\rm Volt}$ and $V_2 = 5.1\;{\rm Volt}$ at $N_{{\rm e}} = 14.5\times 10^6$. The change in the density $\delta n_{{\rm e}}$ due to irradiation was then 
determined from Eq.~(\ref{eqIpv}), it is proportional to the area below the $i_{{\rm pv}}(t)$ curves (shaded region for $V_{{\rm g}} = V_1$). 
Region $(I)$ shows a clear plateau at $n_{{\rm eM}} \simeq 3.5\times 10^{6}{\rm cm^{-2}}$ where the density is independent of the initial density distribution. 
({\bf d}) A magnification of region $(II)$ shows the density in the guard under irradiation, determined approximatively from $n_{{\rm gM}} = (N_{{\rm e}} - n_{{\rm eM}} S_{{\rm d}})/S_{{\rm g}}$.
}
\label{FigIpv}
\end{figure}

We developed the following approach to determine the central density under irradiation $n_{{\rm eM}}$ for different initial densities $n_{{\rm eD}}$.  
We performed compressibility measurements without irradiation as described in Fig.~\ref{FigDensity}
to obtain $n_{{\rm eD}}$ as function of $V_{{\rm g}}$ at a fixed $N_{{\rm e}}$. Then, fixing $V_{{\rm g}}$, we irradiated the electron system with 
on/off pulses of millimeter waves creating a periodic displacement of the electron density $\delta n_{{\rm e}} = n_{{\rm eM}} - n_{{\rm eD}}$. This displacement induces a transient current of image charges on the measuring electrode $i_{{\rm pv}}(t)$, which within a plane-capacitor approximation is related to the change in the electron density $\delta n_{{\rm e}}$ through the relation:
\begin{align}
\delta n_{{\rm e}} = n_{{\rm eM}} - n_{{\rm eD}} = \frac{2}{e \pi R_{{\rm i}}^2} \int i_{{\rm pv}}(t) dt 
\label{eqIpv}
\end{align}
The integral in this equation is evaluated over the time interval where the irradiation is switched off (a derivation of this relation is provided in supplementary note 4) . Combining $\delta n_{{\rm e}}$ with the known values for $n_{{\rm eD}}$, we reconstructed the dependence of $n_{{\rm eM}}$ on the guard potential $V_{{\rm g}}$: the results are shown on Fig.~\ref{FigIpv} for several $N_{{\rm e}}$ values.

Deviations from $n_{{\rm eD}}$ mainly appear in two voltage regions which are in good correspondence with the regions $(I)$ and $(II)$ outlined in Fig.~\ref{FigChiMW}.
In region $(I)$ the density under irradiation exhibits a striking plateau as a function of $V_{{\rm g}}$ with a plateau density 
independent of $N_{{\rm e}}$. We emphasize that this plateau appears because of the cancellation between the decrease of $n_{{\rm eD}}$ at higher $V_{{\rm g}}$ and the increase of the area below $i_{{\rm pv}}(t)$ curves (see Fig.~\ref{FigIpv}.{\bf c} and Eq.~\ref{eqIpv}), it is thus a highly non-trivial experimental result.
These observations confirm the existence of an incompressible phase for electrons on helium 
and appear to exclude an explanation based only on $\sigma_{{\rm xx}} = 0$ since the final state density does not depend on 
the density distribution in equilibrium for a wide range of parameters. 
Instead, our experiments imply the existence of a dynamical mechanism that stabilize the electron density to a fixed value.

We next comment on the sign of $\delta n_{{\rm e}}$. In region $(I)$ the electrons migrate from the guard, where the densities are higher, to the center of the electron cloud, 
whereas in region $(II)$ the trend is opposite and electrons flow from the center towards the edges of the electron cloud. In the latter region narrow density plateaux are also observed however the plateau density value depends on $N_{{\rm e}}$ in contrast to region $(I)$. 
An approximate calculation of $n_{{\rm gM}}$ shown on Fig.~\ref{FigIpv} suggests that in region $(II)$
the transition to an incompressible state occurs owning to the pinning of the density inside the guard region. Thus in the incompressible phase, the electron cloud transfers 
electrons from a high density reservoir region, increasing the density in the low-density regions up to a plateau value. 

{\it Discussion} 

Since the formation of a non-equilibrium density profile increases the electron-electron repulsion energy, 
it is important to estimate the associated energy cost.  
Using the simplified electrostatic model (see supplementary note 5 for a detailed derivation), we find that the energy cost of the redistribution per electron $\Delta_{{\rm e}}$ is approximately:
\begin{align}
\Delta_{{\rm e}} = \frac{e S_{{\rm d}}}{2 \chi_0 N_{{\rm e}}} \delta n_{{\rm e}}^2 
\end{align}
From the experimental values of $\delta n_{{\rm e}} = 1.5\times 10^{6} \;{\rm cm^{-2}}$ and $N_{{\rm e}} = 12.4\times 10^6$
(estimated at point $V_2$ on Fig.~\ref{FigIpv}), we find $\Delta_{{\rm e}} \simeq 0.1\;{\rm eV}$. 
The presence of this large electrostatic barrier, can explain why the incompressible regions occupy a 
narrower $V_{{\rm g}}$ range in the photocurrent data than in the compressibility measurement:
for example region $(I)$ has a width at least 0.6 Volt in Fig.~\ref{FigChiMW} but a width of only 0.3 Volt in Fig.~\ref{FigIpv}.
The main difference between the two techniques is that during compressibility measurements microwaves are always present, maintaining 
the system in a non-equilibrium state, whereas in the photocurrent measurement the microwave on/off pulses continuously reset the system back to its equilibrium state. 
Thus in the photocurrent measurements, the electrons must overcome an increasingly large energy barrier to reach the incompressible state as $V_{{\rm g}}$ increases. 
When the barrier becomes too large, the systems remains in its equilibrium state and the photocurrent vanishes abruptly. 
In contrast, in the compressibility measurement, the electron system remains in the incompressible state as $V_{{\rm g}}$ changes and the electrostatic barrier does not need to be overcome directly, allowing the incompressible state to exist over a wider parameter range. (In the supplementary figures 6-9, we provide a detailed comparison 
between the two measurement techniques and show that they are fully consistent once the described hysteretic behaviour is taken into account).

The estimated charging energy must be provided by the microwave irradiation since it is the only energy source 
in our system, its amplitude corresponds to a surprisingly large number of photons absorbed per electron $\Delta_{{\rm e}}/\hbar \omega \simeq 170$,
particularly in comparison with a two level system that cannot absorb more than one photon.
The energy of the absorbed photons thus needs to be transferred efficiently to other degrees of freedom. It could be transferred by the excitation of higher Landau levels.
However, the theoretical calculations performed by Y. Monarkha to explain the origin of microwave induced resistance oscillations for electrons on helium, suggest that Landau levels higher than the photon energy are unlikely to be strongly populated \cite{Monarkha2011,Monarkha2012}. These calculations give an accurate prediction for the phase of the resistance oscillations allowing us to exclude strong inter-Landau level excitation \cite{Denis3}.

The energy can also be accumulated by transitions within the same manifold of quasi-degenerate Landau levels as they are bent by the confinement potential.
A possible mechanism for this absorption is provided by the negative conductivity models introduced theoretically to explain ZRS in heterostructures \cite{review}. 
It has been predicted that a negative resistance state will stabilize through the formation of domains with a fixed built-in electric field $E_{{\rm c}}$. We performed simulations 
of the electron density profiles for an electric field dependent conductivity model with $\sigma(E) \propto (E^2 - E_{{\rm c}}^2)$ 
which is believed to describe domain formation. Our simulations show that this model tends to fix the electron density gradients in order to generate the built in field $E_{{\rm c}}$. This changes the compressibility of the electron cloud depending on the number of stable domains in the system but without inducing an incompressible state. Thus, new theoretical developments are required to make the domain picture consistent with our experiments.

A recent theoretical proposal suggests that another instability can occur even before the conductivity becomes negative \cite{Entin2013,Entin2014}. 
Due to fluctuations of the microwave electromagnetic field on the wavelength scale $\lambda \simeq 2\;{\rm mm}$ ratchet internal currents are expected to appear under irradiation inside the electron system. The velocity of the induced electron flow will scale as $v \sim \mu_2 E_{\omega}^2$ where $E_\omega$ is the amplitude of the microwave electric field and $\mu_2$ a non-linear response coefficient \cite{Entin2014}. Under steady state conditions this internal flow must be compensated by 
a counter-flow created by internal electric fields with maximal value $e n_{{\rm e}} / \epsilon_0$. 
The balance between the two flows sets a lower bound on the longitudinal mobility that is $\mu_{{\rm xx}}$ stable under irradiation: $\mu_{{\rm xx}} \sim \epsilon_0 \mu_2 E_{\omega}^2/e^2 n_{{\rm e}}$. Mobilities below this threshold are expected to lead to the formation of electron pockets on a scale $\lambda$ with a certain similarly to the stripe/puddles instability occurring in quantum Hall systems \cite{Shkovskii}. 

The above argument predicts the existence of an instability without explaining the pinning of the electron density. A mechanism selecting a particular density value is thus desirable. As a possible mechanism we propose the following scenario. If we assume that a significant fraction of the energy carried by the absorbed photons is transferred into ripplons with a typical wave number given by the inverse magnetic length $k_r \sim \sqrt{e B/\hbar}$ \cite{Lea}, then the helium surface will vibrate at frequency $\omega_r \sim \left( \gamma k_r^3/\rho \right)^{1/2}$ where $\gamma$ and $\rho$ are the helium surface tension and density respectively \cite{monarkha_book}. Provided frequencies are matched, this vibration can interact resonantly with electronic modes. The frequency of the expected surface vibrations is approximately $\omega_r \simeq 2\pi \times 30\;{\rm MHz}$ (for $B = 0.79\;{\rm Tesla}$) and has the same order of magnitude as the low wavelength magneto-shear modes \cite{Golden} which have frequency $\Omega_s \sim \frac{e^2 n_{{\rm e}}^{3/2}}{\epsilon_0 m \omega_{{\rm c}}}$. Equating these frequencies $\omega_r \sim \Omega_s$ we find an equation on the plateau density:
\begin{align}
n_{{\rm e}} \sim \left[ \frac{\gamma}{\rho} \frac{\epsilon_0^2 B^2}{e^2} \left( \frac{e B}{\hbar}  \right)^{3/2} \right]^{1/3} \propto B^{7/6}
\end{align}
For $B = 0.79\;{\rm Tesla}$ this formula gives the right order of magnitude for the plateau density $n_{{\rm e}} \simeq 5.8\times 10^6\;{\rm cm^{-2}}$, 
and the predicted dependence on $B$ is consistent with our observations at different $J = \omega/\omega_{{\rm c}}$ shown on Fig.~\ref{FigChiMW}. 
Even if the described mechanism requires further theoretical studies, it correctly captures  the critical dependence on the electronic density observed in the experiments. For electrons on helium the electronic density mainly controls the strength of electron-electron interactions and the key role 
played by this parameter indicates that the formation of incompressible states is a collective effect involving many electrons. 

In conclusion, we have shown that electrons trapped on a helium surface exhibit incompressible behavior under resonant irradiation conditions corresponding to the formation of a microwave induced zero-resistances state. Their density becomes pinned to a fixed value independent of the applied electrostatic force and on the initial electron distribution profile for a wide range of parameters. The transition to the incompressible state is achieved by overcoming an impressively high electrostatic energy barrier of up to $0.1\;{\rm eV}$ per electron. 
We described the possible energy conversion processes within the electron system that can transform the energy of the absorbed photons into charging energy, and we proposed several competing mechanisms that can render the equilibrium density profile unstable. Since the incompressible behavior emerging from our experiments
is very elegant, we believe that its understanding will stimulate the emergence of new concepts for self-organization in quantum systems.

{\it Acknowledgements:} We are thankful to H. Bouchiat for fruitful discussions and acknowledge support from JSPS KAKENHI Grant Number 24000007. One of us, A.C. acknowledges support from the E. Oppenheimer fellowship and from St. Catharine college in Cambridge. D.K. is supported by internal grant from Okinawa Institute of Science and Technology Graduate University.

{\it Author contributions:} All authors contributed to all aspects of this work.

{\it Methods :} Our experiments were performed in a Leiden dilution refrigerator with base temperature around $25\;{\rm mK}$, the magnetic field was 
provided by a home made superconducting magnet with maximal field of $1\;{\rm Tesla}$. The experimental cell was filled with liquid helium until half 
filling by monitoring the cell capacitance during helium condensation. Electrons were deposited on the helium surface by thermionic-emission from a heated tungsten filament. 
Control of the total electron number $N_{{\rm e}}$ in the cloud was achieved by trapping an initially high concentration of electrons and then lowering the 
confinement voltages allowing excess electrons to escape. The obtained $N_{{\rm e}}$ value was determined from compressibility measurements in equilibrium. 
The frequency of the intersubband transition was tuned in resonance with the photon energy using the linear Starck shift induced by the electric field perpendicular 
to the helium surface which was fixed during our compressibility and photo-current experiments in order to keep the system at intersubband resonance. 
The current probe electrodes were grounded through Stanford Research (SR570) current amplifiers while the potential of the other electrodes was set by Yokogawa DC voltage supplies, this ensure a stable DC voltage on all cell electrodes independently of the irradiation. Numerical simulations were performed by solving the Laplace equations for our cell geometry using a finite elements method \cite{Hecht}.

\clearpage

\section*{A novel incompressible state of a photo-excited electron gas : supplementary materials}

{\bf Supplementary note 1. On the definition of compressibility.} 

Our definition of compressibility deviates from the definition of compressibility in the quantum-Hall effect community since the derivative of the density is taken against a gate voltage and not the chemical potential (position of the Fermi level). Such a definition is not suitable in our case since electrons on helium form a non-degenerate electron gas and the Fermi-energy is much smaller than temperature and is therefore not a relevant parameter, instead we have used the electrostatic potential which is controlled by $V_g$. We note that our experimental technique is otherwise very similar to the capacitive measurements performed on GaAs two-dimensional electron gas to determine the compressibility. For example, except for the dimension of the probe electrode, our measurement geometry is very similar to that in [S.H. Tessmer, P.I. Glicofridis, et. al. ... Nature {\bf 392}, 51 (1998)] even in terms of the non-local excitation scheme used in their experiment. We would also like to emphasize that as $V_g$ decreases the electron gas is progressively compressed towards the center of the electron cloud, which is the expected behavior in a compressibility experiment. Thus we believe that the use of the term compressibility is fully justified.

Finally our definition of compressibility may seem non-local since it involves the derivative against the potential of the guard electrode $V_g$ instead of the potential of the potential of the central electrode $V_d$. Actually only the difference $V_d - V_g$ is important for the electron distribution profile, thus the two definitions are in principle equivalent. Our motivation for choosing $V_g$ as the main control parameter is that we wanted to keep the perpendicular field constant during our measurements, and it is much easier to fix the potential of the top guard electrode $V_{tg}$ ensuring that $V_{tg} - V_{g}$ is constant, than to change the potential of the top central electrodes, which are grounded through the current amplifiers.

{\bf Supplementary Note 2. Derivation of Eq.~(1) in the main text.} We will use $\Delta n_e$ to denote the RMS-oscillation amplitude  of the electronic density below the central electrode of diameter $R_i$ induced by AC voltage with amplitude $V_{ac}$ applied to the bottom and top guard electrodes. The oscillation amplitude $\Delta Q$ of the charge on the top central electrode then can be expressed as: 
\begin{align}
\Delta Q = \frac{e}{2} \Delta n_e \pi R_i^2 
\label{eqDeltaQ}
\end{align}

This equation assumes that electrons are distributed midway between top and bottom electrodes separated by distance $h$. We ensured that this was indeed the case by adjusting the liquid helium level to $h/2$ during the filling of the sample cell using capacitance measurements.
Under these conditions it is strictly valid for an homogeneous charge distribution below the disc for which the density gradients are located at least a screening length $h$ away from the electrode edges. This is the case without microwave irradiation since the density gradients are located at $r \simeq R_d = 1\;{\rm cm} > R_i = 0.7\;{\rm cm}$.   

We next consider the effect of a spontaneous density modulation at a wavenumber $k$ superimposed on the average change in the carrier density on the validity of Eq.~(\ref{eqDeltaQ}). The influence of wavenumbers for which $k h \gg 1$ will be exponentially suppressed by the screening from the electrodes. The effect of wavenumbers satisfying $k R_i \gg 1$ will also be reduced by the averaging over the surface of the measuring electrode. At longer wavelengths $k R_i \ll 1$ the charge density is almost uniform and the error is also expected to be weak. Thus we expect this relation to stay valid for a large range of wave-numbers $k$. 
A possible source of charge density oscillation is the fluctuation of the microwave field on the microwave-wavelength 
scale $\lambda \simeq 2.1\;{\rm mm}$. For this case we have $k h \simeq 7.5$ and $k R_i \simeq 20.5$ we thus expect the effect of the charge modulations to be strongly suppressed as compared to the changes in the average charge density.   

The change in the charge accumulated on the electrodes is related to the current measured with the current amplifier as follows:
\begin{align}
i_{ac} = -\Delta Q \times 2\pi f_{ac}
\label{eqIac}
\end{align}
where $f_{ac}$ is the modulation frequency. This relation assumes a quasi-static limit where the charges follow the external driving. In a lumped-element approximation this is valid provided the modulation frequency $f_{ac}$ is sufficiently low: $2\pi f_{ac} \sigma_{xx}^{-1} C \ll 1$. Here $\sigma_{xx}^{-1}$ is the inverse of the longitudinal conductivity of the electron cloud and $C$ is the coupling capacitance. From our measurements using the Sommer-Tanner technique  without microwaves we know that under our experimental conditions, with typical electron density $\simeq 2\times 10^6{\rm cm^{-2}}$ and magnetic field $B \lesssim 1\;{\rm Tesla}$,  $2\pi f_{ac} \sigma_{xx}^{-1} C \simeq 1$ at a frequency $f_{ac} \simeq 300\;{\rm Hz}$. We performed our compressibility measurements at $f_{ac} = 2\;{\rm Hz}$, thus the inverse conductivity $\sigma_{xx}^{-1}$ of the electron cloud can only affect our measurements when $\sigma_{xx}$ is suppressed by around two orders of magnitude compared with its equilibrium value. As discussed in the main text, the compressibility measurement cannot distinguish between a vanishing conductivity state with $\sigma_{xx} \rightarrow 0$ in which carriers do not respond to the external driving owing to their very long response times and an  "active" mechanism stabilizing the density at a fixed value independent on the applied electrostatic potentials. We note that the photocurrent experiment (Fig.~5 from the main text) allows us to distinguish between these two scenarios by showing that the cloud reaches the same steady state density under irradiation independently on the initial density giving a strong argument in favor of an "active" mechanism. 

Combining equations Eqs.~(\ref{eqDeltaQ}) and (\ref{eqIac}) we find the compressibility as:
\begin{align}
\chi = -\frac{\Delta n_e}{\Delta V_g} = \frac{i_{ac}}{e \pi^2 f_{ac} R_i^2 V_{ac}}
\label{eq1Main}
\end{align}
An implicit assumption here is that $n_e(V_g)$ is a continuous well-behaved function. If $n_e(V_g)$ has a discontinuity  we would theoretically expect a narrow peak smoothed by the lock-in integration-time and instrumental resolution. In the case where an hysteresis loop is present around the discontinuity, the system will stay trapped on the same branch after crossing the discontinuity and the density jump will no longer be correlated with the small voltage modulation $V_{ac}$. In this case the discontinuity may completely disappear from the compressibility traces. In all cases, we do not expect discontinuities to appear as sharp features in the compressibility measurement. 

{\bf Supplementary note 3. Derivation of Eq.~(2) in the main text.} 

We denote $V_e$ the potential of the electron cloud and assumes that the electron cloud extends to both reservoirs, in the plane condensator approximation the charge densities in the central and guard region are:
\begin{align}
n_e &= -\frac{2 \epsilon_0}{e h} \left( 2 V_e - V_d \right) \label{eqne}\\
n_g &= -\frac{2 \epsilon_0}{e h} \left( 2 V_e - V_{tg} - V_g \right) \label{eqng}
\end{align}
In these equations we neglected the contribution of the dielectric constant of liquid helium which is close to one within a few percent error.

Subtracting these two equations and keeping only the AC terms we find :
\begin{align}
\Delta n_e - \Delta n_g = - \frac{4 \epsilon_0}{e h} V_{ac}
\end{align}
Here we remind readers that the AC potential is applied to both the top and bottom guard electrodes.

Charge conservation in the cloud leads to:
\begin{align}
S_d \Delta n_e + S_g \Delta n_g = 0
\end{align}

where $S_d$ and $S_g$ are the surface areas of the central and guard electrodes respectively. We thus obtain the compressibility defined in the text as:
\begin{align}
\chi_0 = -\frac{d n_e}{d V_g} = -\frac{\Delta n_e}{V_{ac}} = \frac{4 \epsilon_0}{e h} \frac{1}{1+S_d/S_g}
\label{defchi}
\end{align}

For reference, note that we can deduce the potential of the electron cloud with a given total number of electrons $N_e$ by inserting Eqs.~(\ref{eqne}),(\ref{eqng}) into:
\begin{align}
S_d n_e + S_g n_g = N_e,
\end{align}
this leads to:
\begin{align}
V_e = \frac{V_d S_d + S_g (V_g + V_{tg})}{2 (S_d + S_g)} - \frac{e h N_e}{4 \epsilon_0 (S_d + S_g)}
\end{align}

Another way of deriving these equations is by minimization of the charging energy, this derivation also allows us to treat naturally the case where one of the reservoirs is fully depleted under the action of the gate potentials.

{\bf Supplementary note 4. Derivation of Eq.~(3) in the main text.} 

This equation connects the change in the electron density under irradiation and the photo-current : 
\begin{align}
\delta n_e = n_{eM} - n_{eD} = \frac{2}{e \pi R_i^2} \int i_{pv}(t) dt 
\label{ipv}
\end{align}
where integration is performed over the time interval in which the irradiation is switched off.
This equation is derived by combining Eq.~(\ref{eqDeltaQ}) with $i_{pv}(t) = -\Delta Q/\Delta t$ where the ratio is taken in the sense of a time derivative.

The assumptions behind this equation are in practice slightly different from the assumptions behind Eq.~(\ref{eq1Main}), indeed since integration is performed over the off phase of the microwave cycle the dynamics of the electrons is known from the dark state properties and the only unknown is the state under microwave irradiation from which relaxation back to equilibrium starts. We thus expect this relation to be less sensitive to short wave-length fluctuations since they will quickly relax back to a homogeneous density without irradiation. Since the two relations correspond to slightly different assumptions on the unknown steady state under irradiation, it is important to ensure that they produce consistent results (this is shown in Figs. ~\ref{FigCR1},\ref{FigChi07},\ref{FigChiNeg})

{\bf Supplementary note 5. Derivation of Eq.~(4) in the main text.} 

We assume a more general configuration with fixed $n_e$ and $n_g$. The electrostatic potentials in the center $V_{ed}$ and in the guard $V_{eg}$ are then different but their values are still given by relations similar to Eqs.~(\ref{eqne},\ref{eqng}): 
\begin{align}
V_{ed} = \frac{1}{2} \left( V_d -\frac{e h}{2 \epsilon_0} n_e \right) \label{eqne2}\\
V_{eg} = \frac{1}{2} \left( V_{tg} + V_g - -\frac{e h}{2 \epsilon_0} n_g \right) \label{eqng2}
\end{align}

The electrostatic energy $U(n_e, n_g)$ of the system in a plane capacitance approximation is then:
\begin{align}
U(n_e, n_g) &= \frac{\epsilon_0}{h} \left[ S_d V_{ed}^2 + S_d (V_{ed} - V_d)^2 + S_g (V_{eg} - V_g)^2 + S_g (V_{eg} - V_{tg})  \right]
\label{eqUne}
\end{align}

The densities $n_e$ and $n_g$ are related by charge conservation $S_d n_e + S_g n_g = N_e$, the electrostatic energy $U$ can thus be viewed as a function of $n_e$ alone $U = U(n_e)$. Moreover an inspection of Eqs.(\ref{eqne2},\ref{eqng2},\ref{eqUne}) shows that $U(n_e)$ is a second order polynomial in $n_e$, for our purposes we will need only the highest order term: 
\begin{align}
U(n_e) &= \frac{e^2 h}{8 \epsilon_0} n_e^2 S_d \left[ 1 + \frac{S_d}{S_g} \right] + O(n_e)\\
&= \frac{e S_d}{2 \chi_0} n_e^2 + O(n_e)
\end{align}
where we remind that $\chi_0 =  \frac{4 \epsilon_0}{e h} \frac{1}{1+S_d/S_g}$. Since the electrostatic energy is minimal for the dark electron density, we obtain finally :
\begin{align}
  U(n_e) = \frac{e S_d}{2 \chi_0} (n_e - n_{eD})^2 + \rm{const}
\end{align}
where the constant is independent on $n_e$.

Dividing by the total electron number, and using the notation $\delta n_e = n_{eM} - n_{eD}$ we obtain the expression for the energy cost of the redistribution per electron which was given without derivation in the main article:
\begin{align}
\Delta_e = \frac{U(n_{eM}) - U(n_{eD})}{N_e} =  \frac{e S_d}{2 \chi_0 N_e} \delta n_e^2 
\end{align}

\clearpage 

{\bf Supplementary figure \ref{FigMWPow}: Dependence on microwave power}

As demonstrated in Supplementary Figure~\ref{FigMWPow} the microwave power  changes the surface of the incompressible regions on the $(n_{eD}, n_{gD})$ plane without changing their horizontal/vertical boundaries. The observation that the $n_{eD}$ and $n_{gD}$ density boundaries are independent on the microwave power supports our idea of a resonance that occurs around certain electron densities independently of the microwave power. At lowest microwave power the narrow incompressible region clearly follows lines of constant $N_e$.
This can be understood from the hysteretic behavior which we described in the main manuscript. At lowest microwave powers, the incompressible state can only form within a narrow range of $N_e$ values. Once the incompressible state is formed, it can exist as a meta-stable state over a wide range of guard voltages; which when converted into the corresponding dark density parameters $n_{gD}$ and $n_{eD}$ give a narrow incompressible stripe tilted along the slope $N_e = {\rm const}$. 

\begin{figure}[h]
  \centerline{\includegraphics[clip=true,width=16cm]{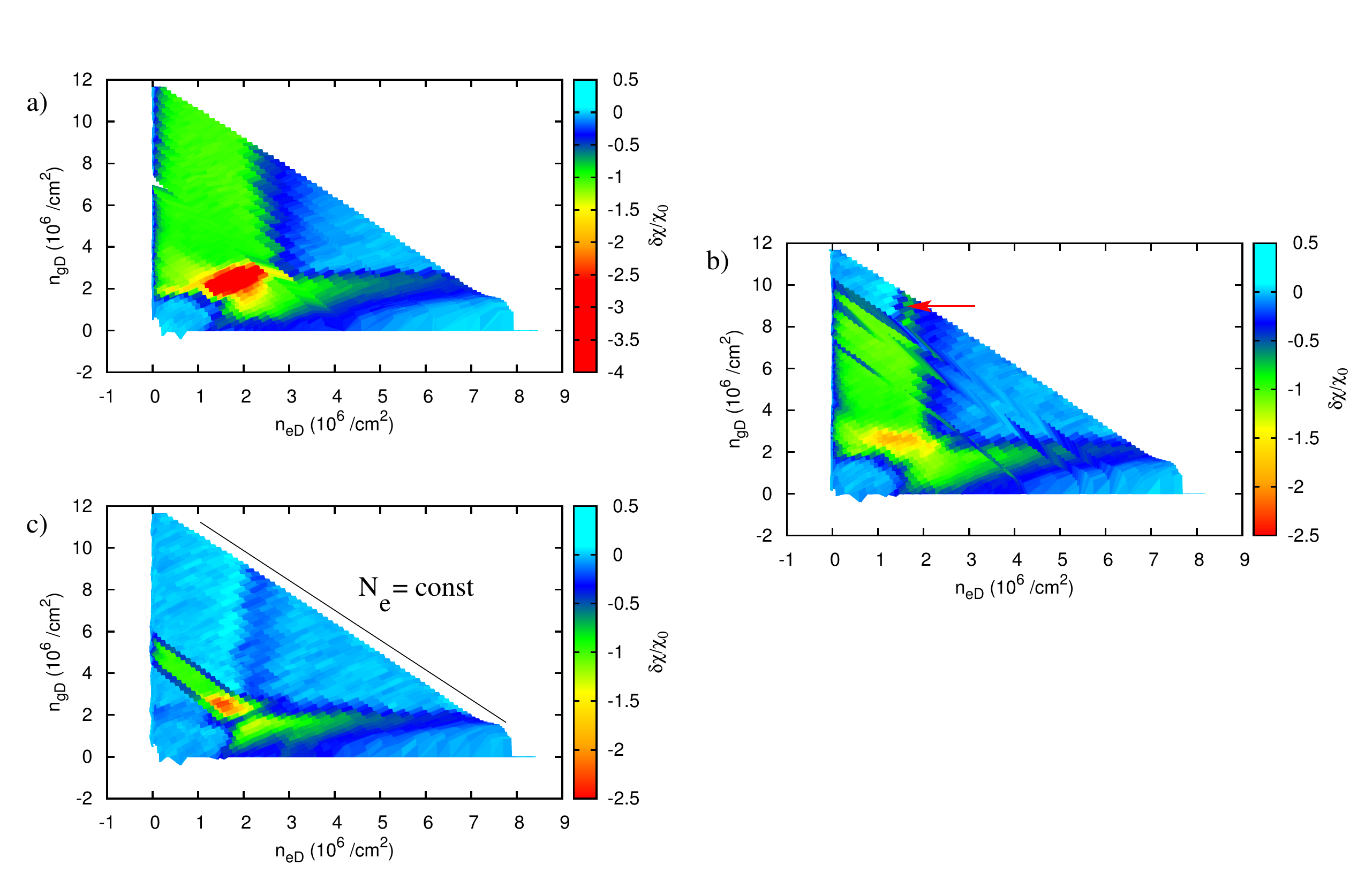}}
\caption{Evolution of the incompressible regions (colored in green) as a function of the dark densities $n_{eD}$ and $n_{gD}$ for different microwave powers at $J = 6.25$. Panel a) corresponds to the maximal microwave power and is identical to Fig.~4 in the main manuscript but with a different color-code. In panel b) the microwave power was reduced to half of its maximal value, and in panel c) the microwave power was $30\%$ of the maximum. The red-arrow on panel b) indicates the peak in $\delta \chi/\chi_0$ just outside the incompressible region supporting our suggestion of a density dependent resonance. The explicit dependence on $N_e$ supports the existence of long-range correlations in the system since not only the local densities are important.
}
\label{FigMWPow}
\end{figure}

\clearpage 

\begin{figure}[h]
  \centerline{\includegraphics[clip=true,width=12cm]{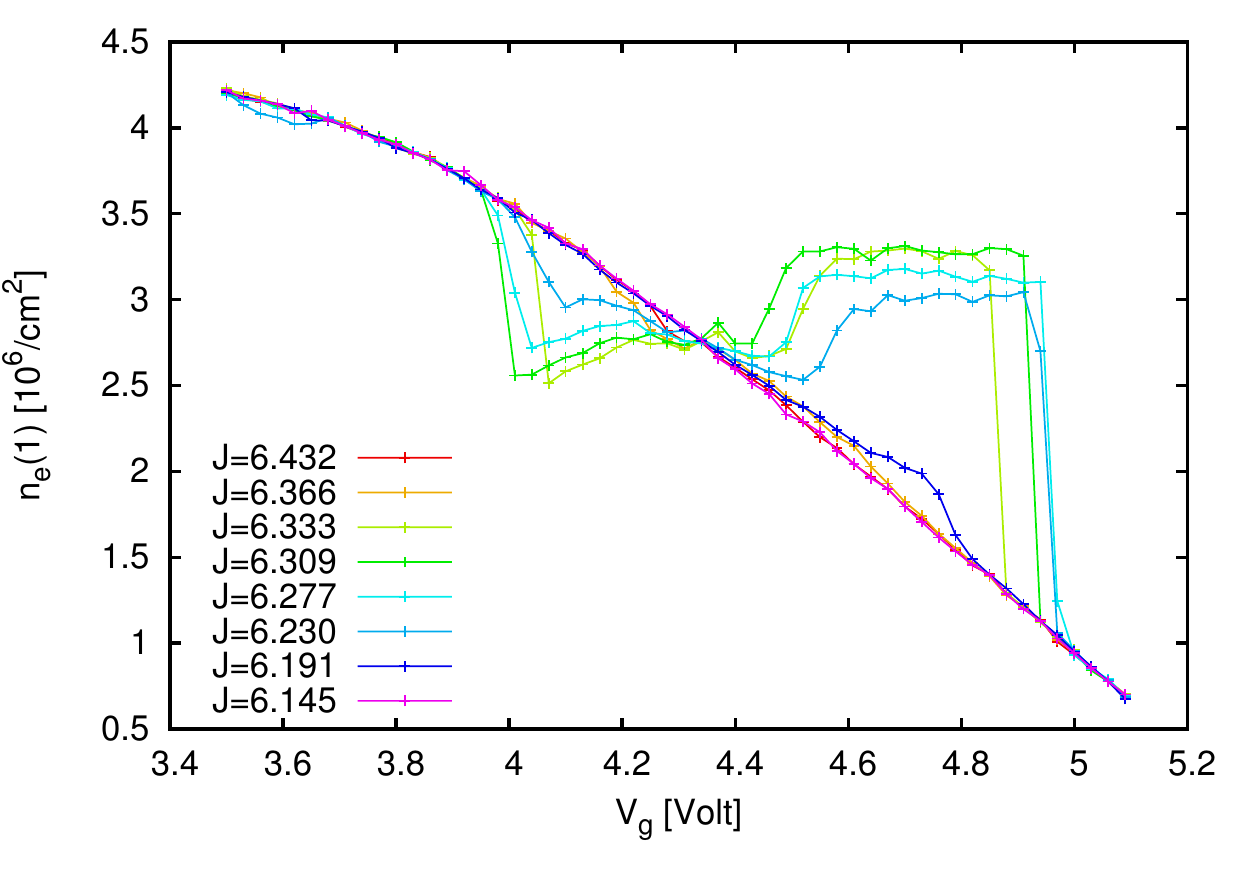}}
\caption{Density under microwave irradiation $n_{eM}$ obtained from the photo-current technique as a function of $V_g$ for different values of $J = \omega/\omega_c$ around $J \simeq 6.25$. The experimental protocol is identical to Fig.~5 from the main manuscript. The total number of electrons in the cloud during this experiment was estimated to be: $N_e \simeq 11.5 \times 10^{6}$. Incompressible plateaux only appear for $J \in (6.23, 6.333)$, outside this interval values very close to $n_{eD}$ are recovered.
}
\label{FigJzoom}
\end{figure}

{\bf Supplementary Figure \ref{FigJzoom}. Magnetic field dependence around $J = n+1/4$ ($n$ integer).}

We did not observe any strong dependence on the parameter $J = \omega/\omega_c$  as long as its value was near a minimum of the MIRO oscillations (integer plus 1/4 offset), the corresponding data is shown in supplementary figure~\ref{FigJzoom}. When the deviation of $J$ became too large the incompressible states disappeared abruptly and the equilibrium dependence $n_e(V_g)$ was recovered. This shows that the integer part of $J$ is the important parameter for the study of the magnetic field dependence of the incompressible states.

\clearpage

\begin{figure}[h]
  \centerline{\includegraphics[clip=true,width=12cm]{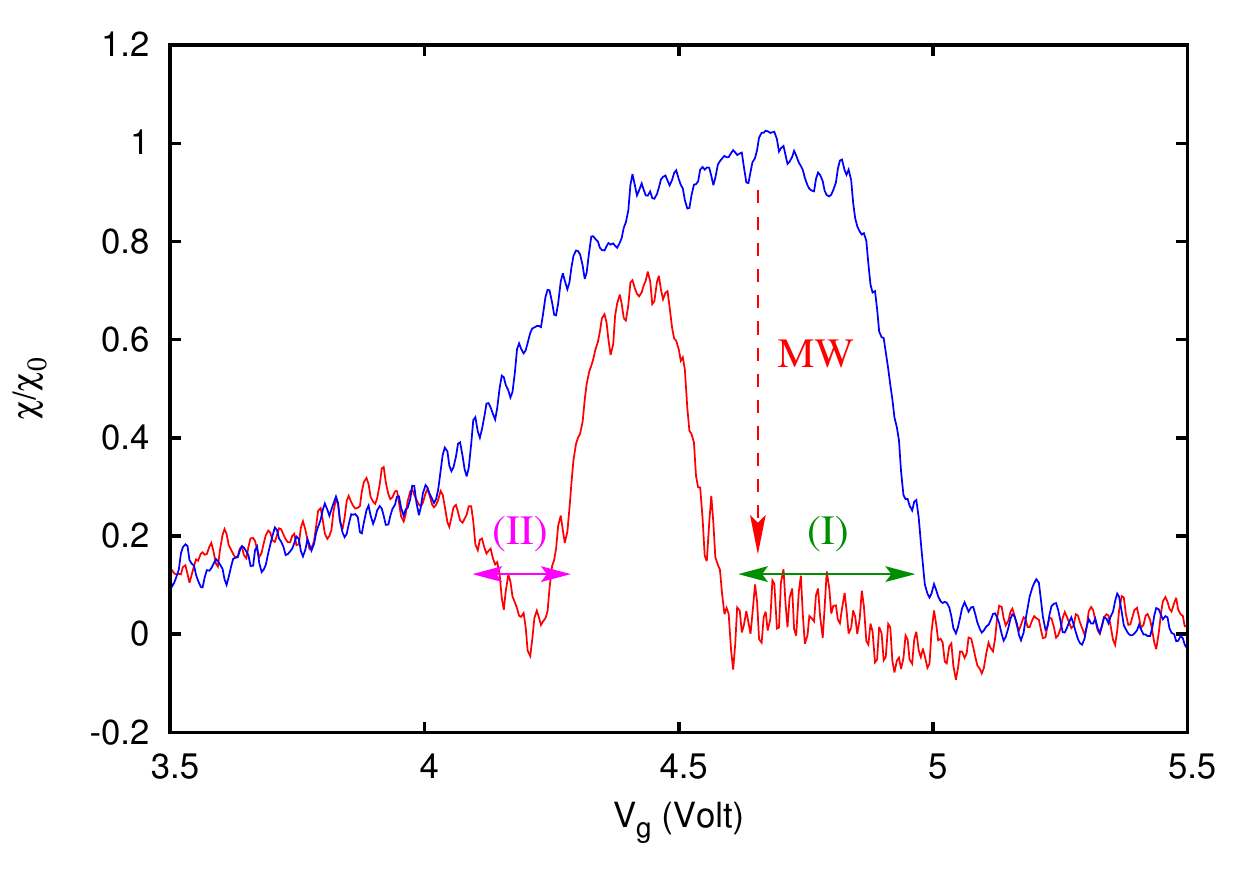}}
\caption{Compressibility in equilibrium (blue) and under microwave irradiation (red) for $N_e = 7.2 \times 10^6$ at $J = 10.25$. The structure is very similar to Fig.~4 (main text) with the appearance of two distinct incompressible regions.
}
\label{FigChiJ10.25}
\end{figure}

{\bf Supplementary Figure~\ref{FigChiJ10.25}: Compressibility at $J = 10.25$}
 
\clearpage 

\begin{figure}[h]
  \centerline{\includegraphics[clip=true,width=12cm]{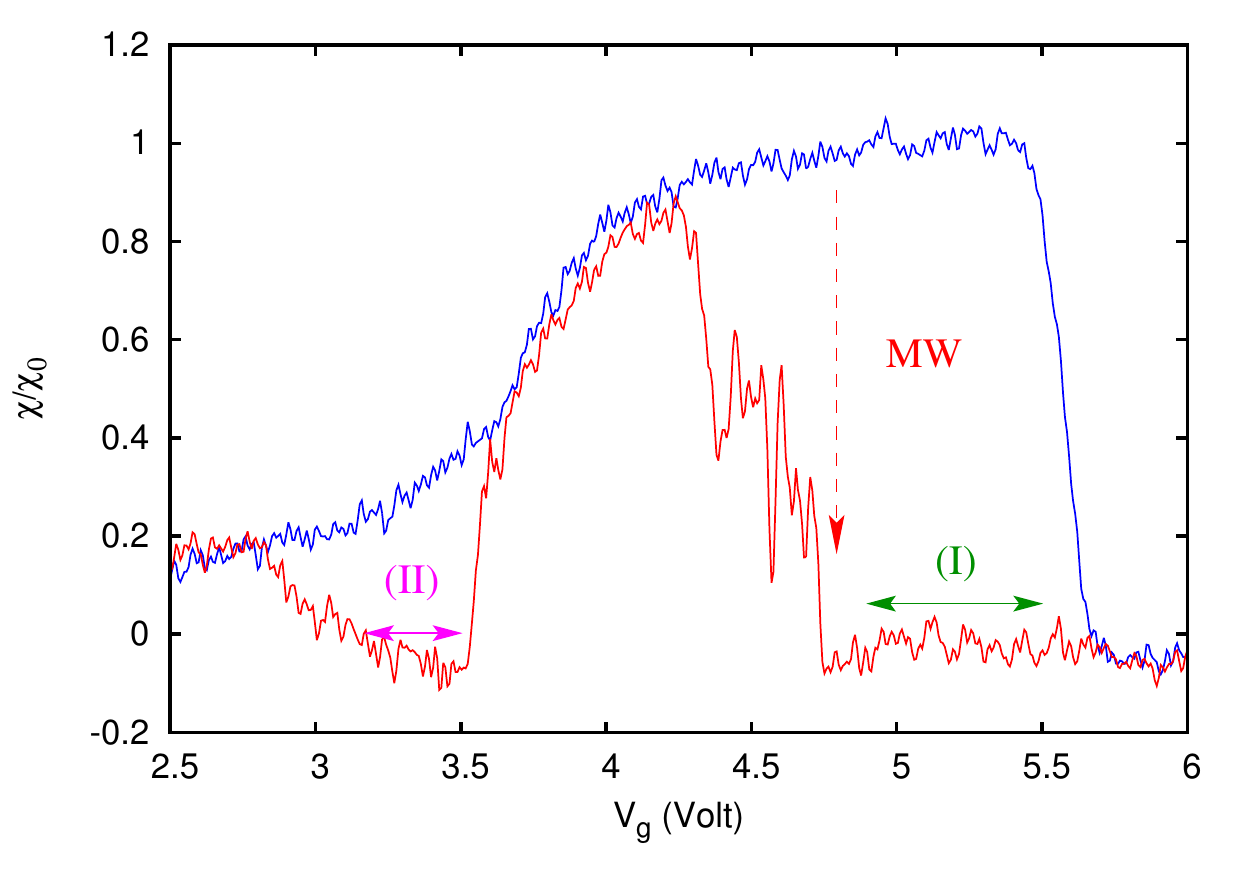}}
\caption{Compressibility in equilibrium (blue) and under microwave irradiation (red) for $N_e = 14 \times 10^6$ at $J = 5.25$. The structure is very similar to Fig.~4 (main text) with the appearance of two distinct incompressible regions.
}
\label{FigChiJ5.25}
\end{figure}

{\bf Supplementary Figure~\ref{FigChiJ5.25}: Compressibility at $J = 5.25$}
 
\begin{figure}[h]
  \centerline{\includegraphics[clip=true,width=12cm]{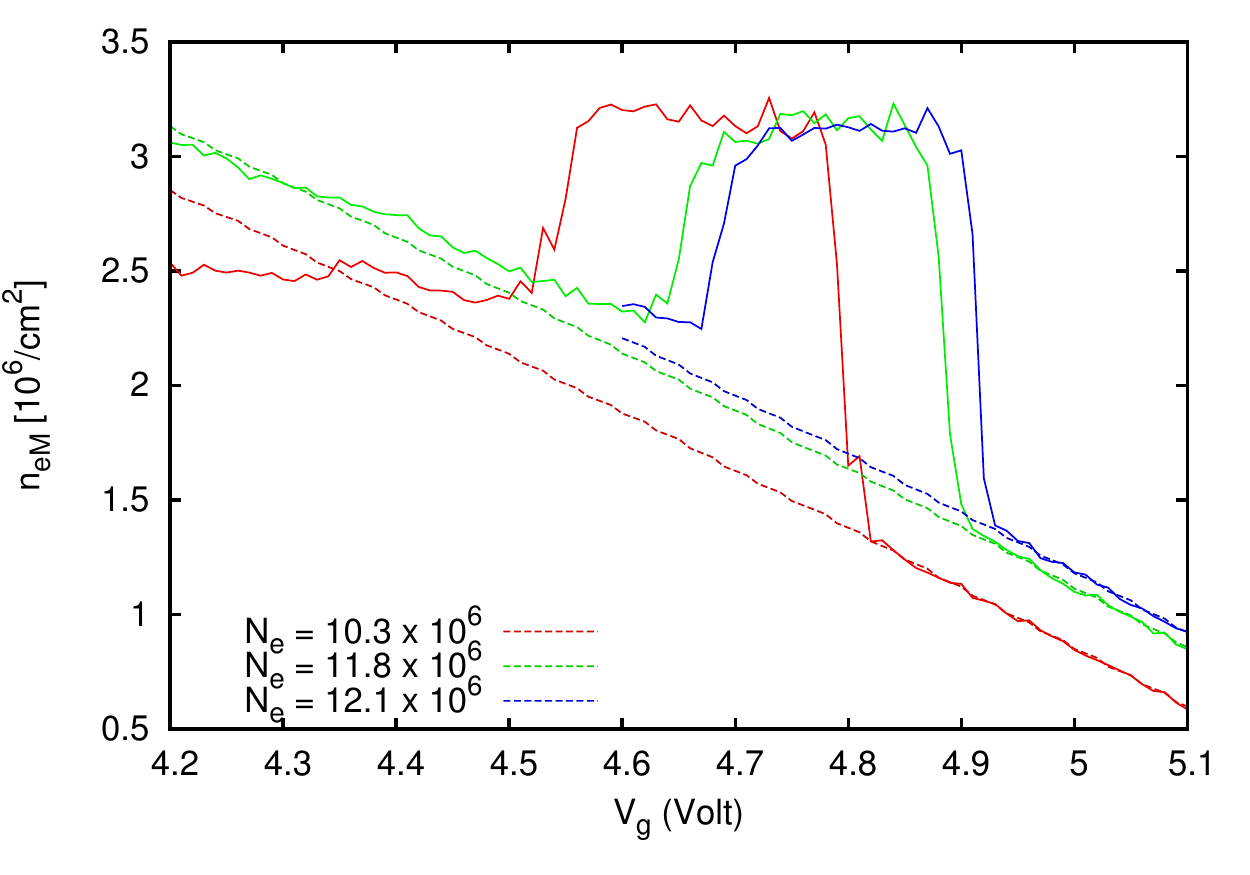}}
\caption{Density $n_{eM}$ under microwave irradiation as function of the guard voltage $V_g$ obtained from the photocurrent technique at $J = 5.25$ for different $N_e$ values. This figure is similar to  Fig.~5 in the main text plotted for $J = 6.25$ (solid lines and also showing the appearance of density plateaux under irradiation). Solid and dashed lines represent $n_{eM}$ and $n_{eD}$ respectively.
}
\label{FigIpv5.25}
\end{figure}

\begin{figure}[h]
  \centerline{\includegraphics[clip=true,width=12cm]{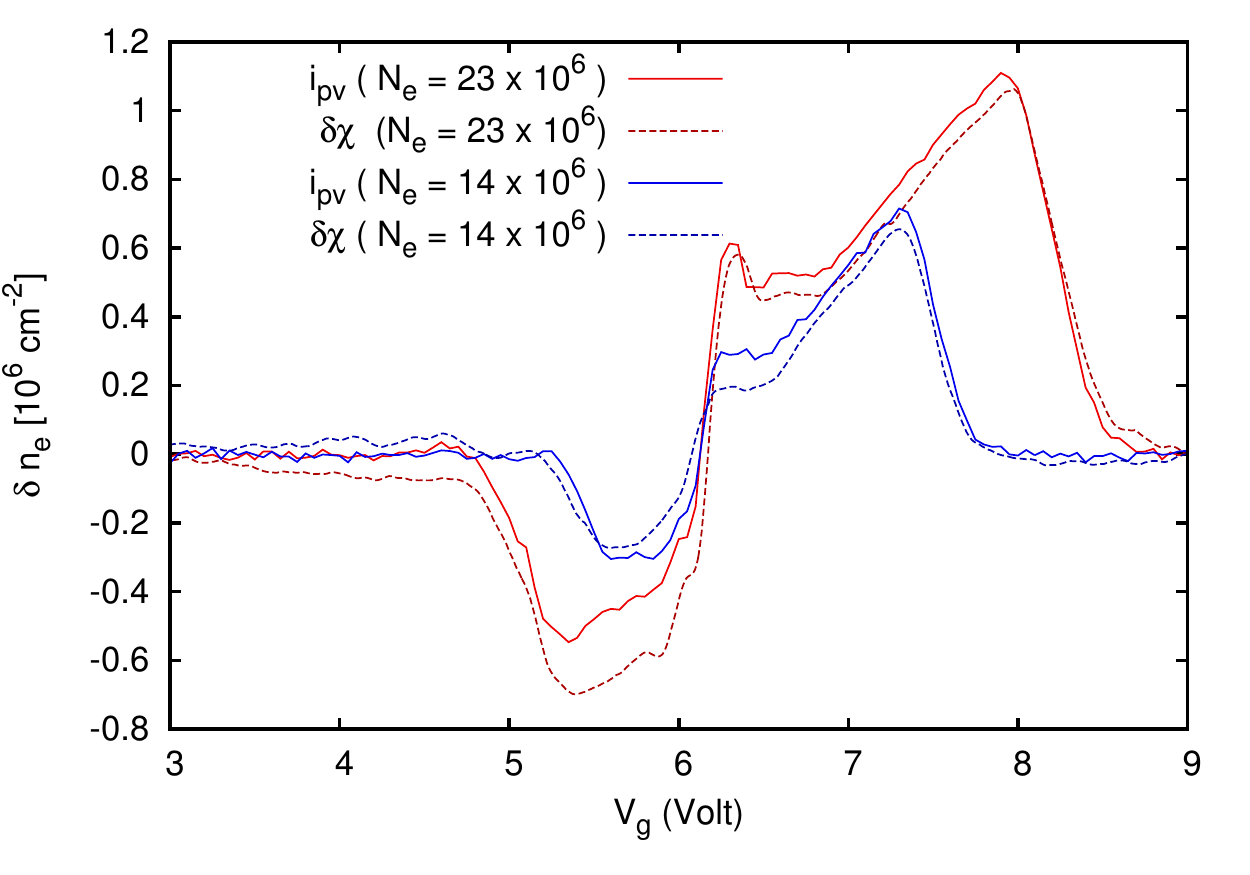}}
\caption{Change in the electron density  $\delta n_e$  under cyclotron resonance conditions measured through the photocurrent (solid curves obtained from Eq.~(\ref{eq:dnepv})) and compressibility techniques (dashed curves obtained from Eq.~(\ref{eq:dnechi})). Cyclotron resonance was excited at a frequency of $\omega = 2 \pi \times 14\;{\rm GHz}$ corresponding to $B = 0.5\;{\rm Tesla}$, a stronger holding field was applied compared to the case of intersubband excitation $V_d = V_{g} - V_{tg} = 6\;{\rm Volt}$. The procedure is otherwise identical to that described in the main text in Fig.~4 and 5 in the main text for experiments at intersubband resonance.
}
\label{FigCR1}
\end{figure}

\clearpage 
{\bf Supplementary Figure~\ref{FigIpv5.25}: Density from the photocurrent technique at $J = 5.25$}

\clearpage

\begin{figure}[h]
  \centerline{\includegraphics[clip=true,width=12cm]{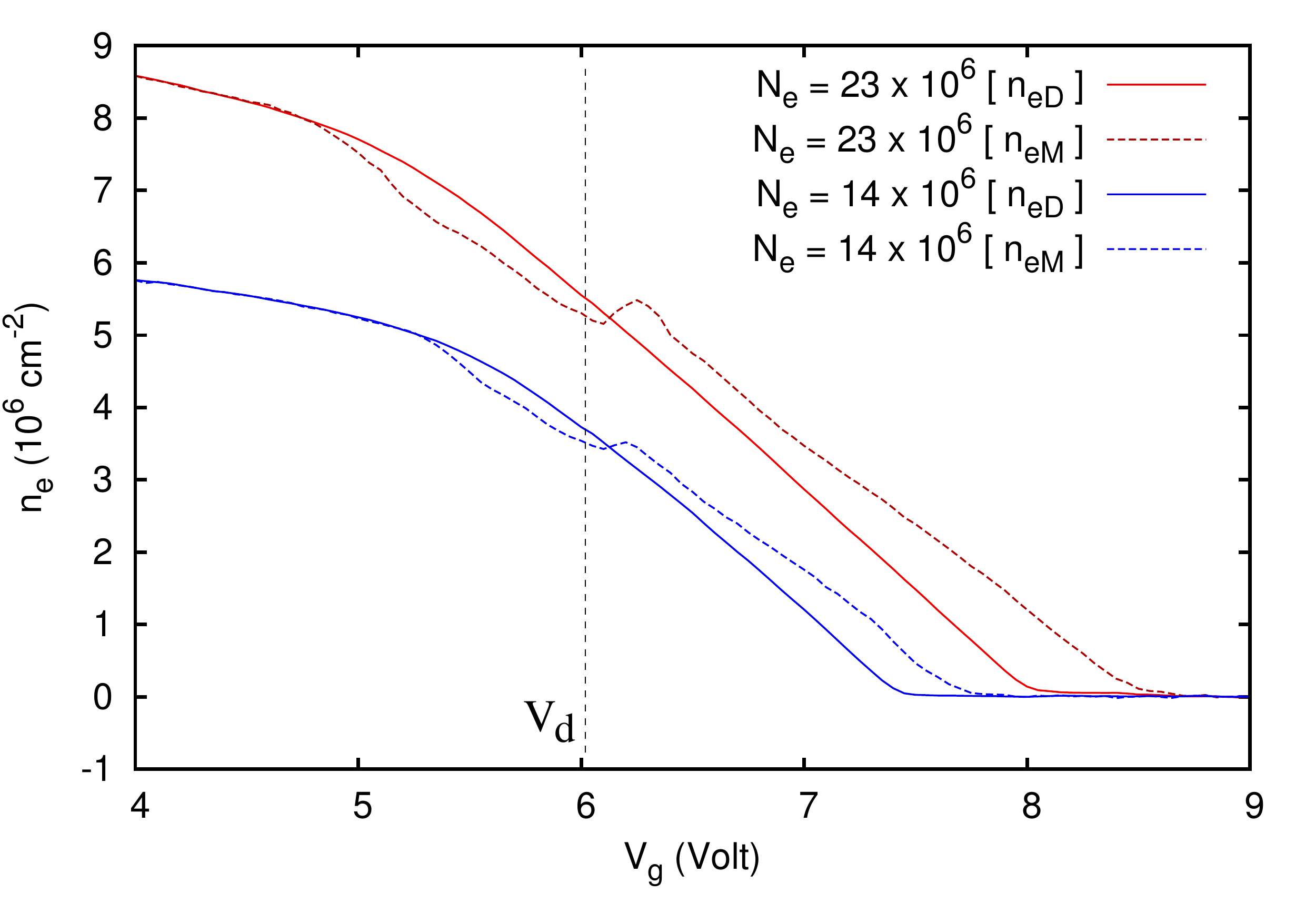}}
\caption{Electron density in the dark (solid lines show $n_{eD}$) and under cyclotron resonance conditions (dashed lines for $n_{eM}$) measured by combining photocurrent and dark compressibility measurements as in Fig.~5 from the main text. Density plateaux are not  observed in this case even if there is a substantial change in the electronic density. The parameters are the same as those in Fig.~\ref{FigCR1}. The shape of the observed dependence is consistent with heating-induced broadening of the dark $n_{eD}(V_g)$ dependence.  However, the presence of an anomaly around $V_g = V_d$ suggests that other physical effects may also be important.
}
\label{FigCR2}
\end{figure}

\begin{figure}[h]
  \centerline{\includegraphics[clip=true,width=16cm]{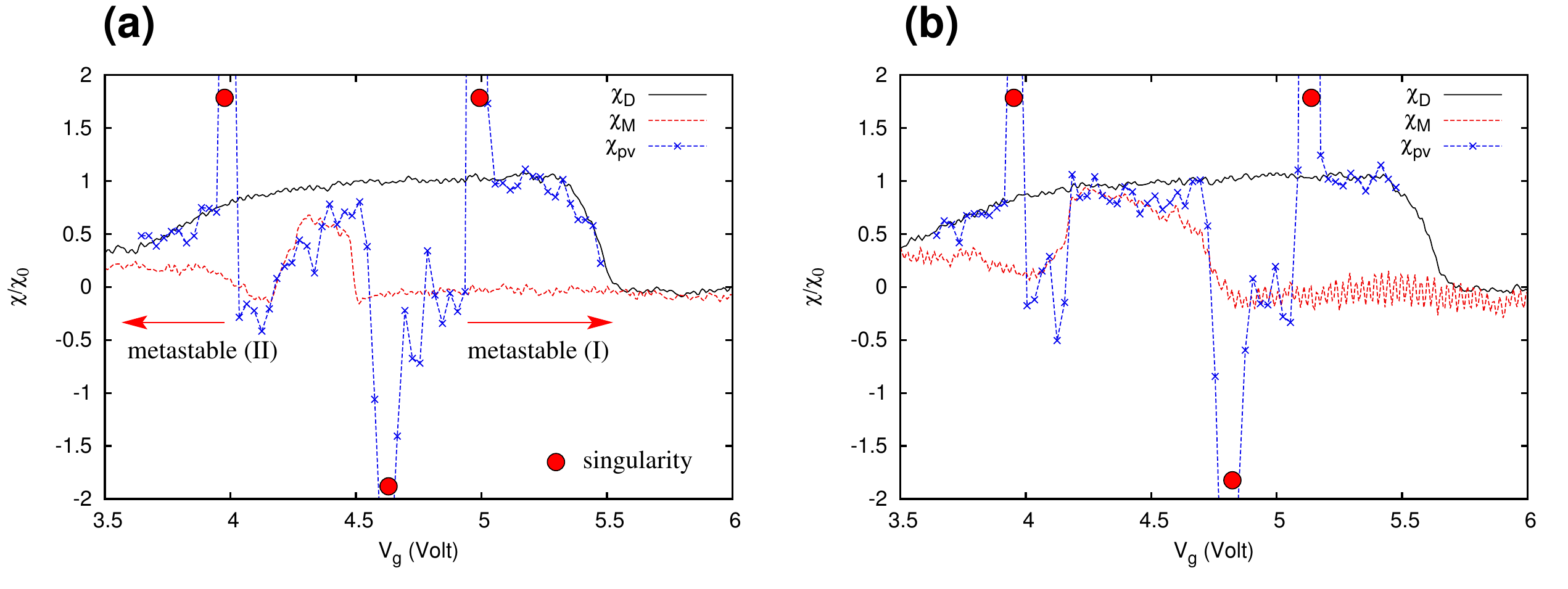}}
\caption{The dark and red curves  respectively show $\chi_D$ and $\chi_M$ measured with the low-frequency lock-in technique for $N_e = 12.4 \times 10^6$ ({\bf a}) and $N_e = 14.5 \times 10^6$ ({\bf b}). The blue curve shows $\chi_{pv}$ calculated by numerical differentiation of the electron density $n_{eM}$ obtained from photocurrent measurements (as a consequence of the differentiation procedure the blue trace is more noisy than the lock-in measurements). The data shown in this figure is taken from the data shown in Figs.~4 and 5 in the main text. 
}
\label{FigChi07}
\end{figure}

\begin{figure}[h]
  \centerline{\includegraphics[clip=true,width=16cm]{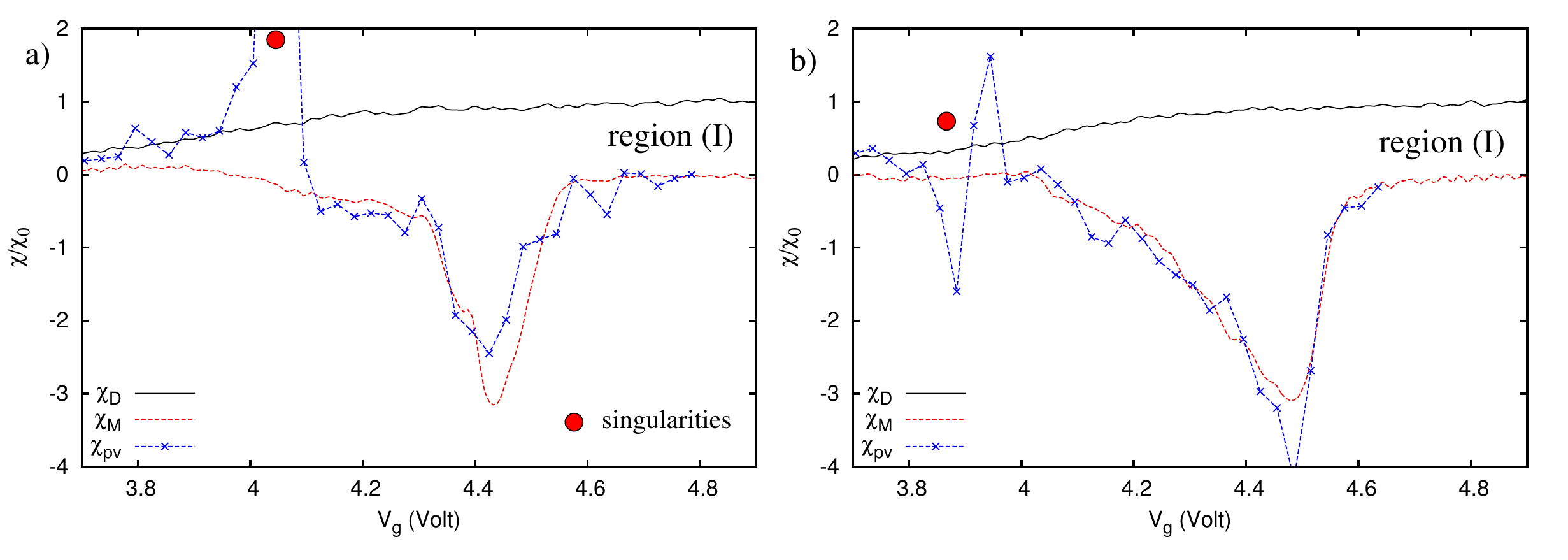}}
\caption{Comparison between $\chi_D$, $\chi_M$ and $\chi_{pv}$ measured at $N_e = 9 \times 10^6$ (left) and $N_e = 5.5\times 10^6$ (right) at $J = 6.25$. This data corresponds to $V_g$ scans across the negative compressibility region shown in Fig.~4 (main text).
}
\label{FigChiNeg}
\end{figure}

{\bf Supplementary Figures~\ref{FigCR1} and \ref{FigCR2}. Consistency between compressibility and photo-current measurements at cyclotron resonance.}

The compressibility under microwave irradiation and the photo-current measurements provide two independent ways of probing the density of the system under microwave irradiation and it is important to verify their consistency. We expect the following relation to hold between the two measurements: 
\begin{align}
\delta n_e = n_{eM} - n_{eD} &= \frac{2}{e \pi R_i^2} \int_{Off} i_{pv}(t) dt \label{eq:dnepv} \\
&= \int_{V_g}^{\infty} \delta \chi \; dV  \label{eq:dnechi} 
\end{align}
Here Eq.~(\ref{eq:dnepv}) reproduces Eq.~(3) from the main text and Eq.~(\ref{eq:dnechi}), which follows from the integration of $\delta \chi = \chi_M - \chi_D$ that is the difference between the compressibility $\chi = -\frac{d n_e}{d V_g}$ under irradiation and in the dark. 

To demonstrate that these two independent measurements indeed lead to the same values of $\delta n_e$ we performed experiments under cyclotron resonance conditions. As shown recently cyclotron irradiation can significantly heat the electron system inducing considerable changes in the electron density [ A. O. Badrutdinov, L. V. Abdurakhimov, and D. Konstantinov  {\it  Cyclotron resonant photoresponse of a multisubband two-dimensional electron system on liquid helium}, Phys. Rev. B. {\bf 90}, 075305 (2014) ]. This provides us with a model case to confirm the consistency between Eqs.~(\ref{eq:dnepv}) and (\ref{eq:dnechi}) in a regime where the mechanism driving the electron redistribution appears to be better understood. The results of our experiment are presented in Fig.~\ref{FigCR1} and show a good agreement between the two techniques. Fig.~\ref{FigCR2} shows the evolution of the density under irradiation as a function of $V_g$.
Although the change in the electron density  $\delta n_e \simeq 1\times 10^{6}\;{\rm cm^{-2}}$ is comparable, but smaller than the changes reported in the main text. Moreover, the density plateaux do not form under cyclotron resonance conditions. 

\clearpage 

{\bf Supplementary Figures~\ref{FigChi07} and \ref{FigChiNeg}. Consistency between compressibility and photo-current measurements in incompressible phases.}

In supplementary Figs.~\ref{FigCR1} and \ref{FigCR2}, using the example of cyclotron resonance, we showed that in general the compressibility and photo-current measurements should indeed be consistent. We also verified this for the case of intersubband resonance in regimes where only a small change in the compressibility was observed without leading to the formation of incompressible phases. However, when incompressible states form, the two techniques do not coincide perfectly. For example the voltage span of the incompressible regions is much larger in the compressibility data than in the photocurrent data. Indeed, as seen from Figs.~4 and 5 in the main text region $(I)$ occupies the voltage range $V_g \in (4 \;{\rm Volt}, 5 \;{\rm Volt})$ according to the compressibility data, but the width of the range is only $0.25\;{\rm Volt}$ according to the photocurrent data for the same number of electrons.

To clarify the origin of this discrepancy we computed the compressibility by numerically differentiating  the electron density under irradiation as obtained 
from the photocurrent measurement; we denote $\chi_{pv}$ as the corresponding value. In Fig.~\ref{FigChi07} we compare $\chi_{pv}$ with the compressibilities $\chi_D$ and $\chi_M$ measured with the lock-in technique in the dark and under irradiation. In this comparison we used the data shown in the main text at $J = 6.25$ at $N_e = 12.5 \times 10^6$ and $N_e = 14.5 \times 10^6$. Except at the singular points that appear due to the differentiation of the abrupt features in Fig.~5 (main text), $\chi_{pv}$ follows the values of $\chi_{D}$ or $\chi_{M}$ switching between the two curves at the singular points located at approximately $V_g \simeq 4$ and $\simeq 5\;{\rm Volt}$.

This suggests, as described in the main text, that at the switching points the electrostatic energy barrier becomes too high to allow a direct transition to the incompressible state from the dark electron density, collapsing the compressibility $\chi_{pv}$ onto the dark branch $\chi_D$. For $\chi_M$ the incompressible behavior can extend over a wider range.
Indeed once the system has formed an incompressible state a small change in $V_g$ does not lead to the formation of a large energy barrier. The incompressible state can thus continue to exist as a metastable state that cannot be reached directly from the dark electron density distribution. In this sense, the formation of the incompressible state is strongly hysteretic. 

The compressibility measurements under irradiation also show the existence of a region with $\chi_M < 0$ around $n_{eD} \simeq n_{gD} \simeq 2\times 10^6\;{\rm cm^{-2}}$. 
We confirmed that similar negative compressibility values can be obtained independently from the numerical differentiation of the photo-current. 
The corresponding data is shown on Fig.~\ref{FigChiNeg}. Inside the negative-compressibility region electrons are repelled from the more positive potential 
effectively behaving as carriers with a positive charge. These experiments show that our two measurement techniques remain consistent even in this 
exotic state of the electron cloud, highlighting that the discrepancies are indeed due to the presence of singular points in $n_e(V_g)$ dependence under irradiation. 
We have not focussed on this regime in the main text since it appears when incompressible region $(I)$ with fixed $n_{eM}$ 
and region $(II)$ with fixed $n_{gM}$ merge into the same $V_g$ range. It is thus a consequence of the interaction between 
the two regions and corresponds to a more complex regime in which the electron system cannot reach a stable final state by increasing the density in one of the reservoirs.

\clearpage 

\begin{figure}[h]
  \centerline{\includegraphics[clip=true,width=14cm]{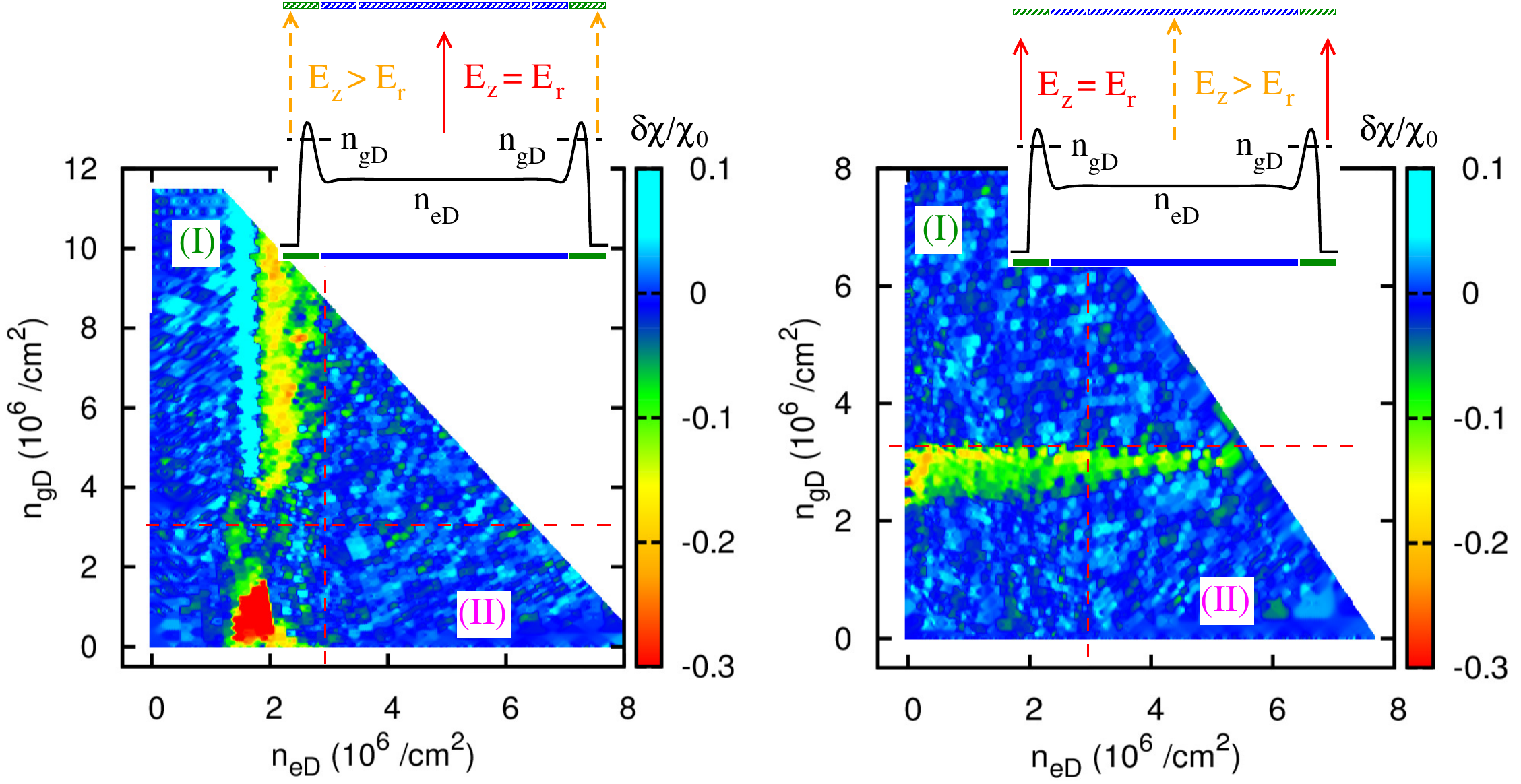}}
\caption{Change of the compressibility  $\delta \chi = \chi_M - \chi_D$  under microwave irradiation as a function of equilibrium density in the center and in the guard under different irradiation conditions. In the left panel, the perpendicular field in the guard region is detuned from the intersubband resonance by setting $V_{g} - V_{tg} = 5.24\;{\rm V}$, while the central region is kept in resonance setting $V_d = 4.24\;{\rm V}$ as previously. 
Under these conditions a weak change in the compressibility $|\delta \chi|/\chi_0 \le 0.2$ appears in the density window bounded by $n_{eD} \le 3 \times 10^6\;{\rm cm^{-2}}$ and incompressible behavior is observed only when the guard is almost depleted $n_{gD} \le 2 \times 10^6\;{\rm cm^{-2}}$.
In the experiment whose results are shown on the right panel, the center of the cell is detuned from the resonance $V_b = 5.24\;{\rm V}$. In this case $\delta \chi$ is non zero along the line $n_{gD} \simeq 3 \times 10^6\;{\rm cm^{-2}}$. 
}
\label{FigChiLoc}
\end{figure}

\begin{figure}[h]
  \centerline{\includegraphics[clip=true,width=12cm]{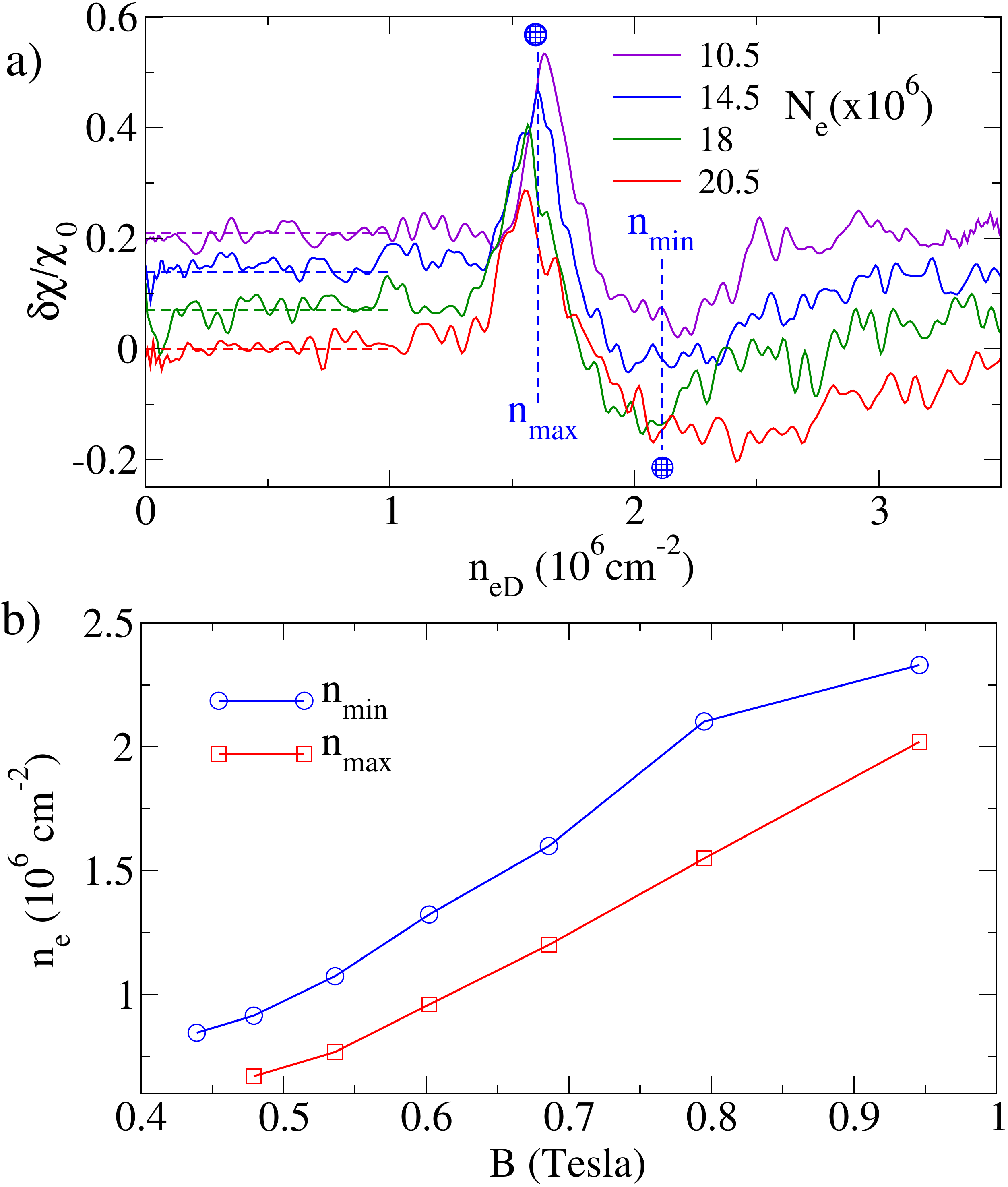}}
\caption{Panel {\bf a}) shows $\delta \chi/\chi_0$ as function of $n_{eD}$ at different $N_e$ using the data from Fig.~\ref{FigChiLoc}, a narrow maximum is observed at a characteristic density $n_{eD} = n_{max} \simeq 1.6\times 10^6\;{\rm cm^{-2}}$ and a broader minimum is observed at $n_{eD} = n_{min} \simeq 2.1\times 10^6\;{\rm cm^{-2}}$ (maximum and minimum values taken for $14.5\times 10^6{\rm cm^{-2}}$). The position of the minima/maxima are almost independent on the boundary conditions which are controlled by the total number $N_e$ of electrons, indeed $n_{max}$ changes only by around $5\%$ when $N_e$ is changed by a factor of two. The presence of those peaks, specially of the narrow maximum, are indicative of a resonance which appears around particular values of the local electron density. We believe that this resonance, which we tentatively attributed to a resonant interaction between plasmons and ripplons is a precursor to the formation of the incompressible state in the entire system when all the system is excited at intersubband resonance. The curves are offset vertically for clarity, the origin for each curve is shown by the horizontal dashed lines. Figure {\bf b}) shows the evolution of the densities $n_{max}$ and $n_{min}$ for different magnetic fields corresponding to different fractions $J = \omega/\omega_c = n+1/4$ (where $n$ is an integer) measured at $N_e = 14.5\times 10^6$, an approximately linear dependence is observed which is consistent with the scaling proposed in Eq.~(5) in the main text.
}
\label{FigDosLoc}
\end{figure}

{\bf Supplementary Figures~\ref{FigChiLoc} and \ref{FigDosLoc}. Detuning of the perpendicular electric field.}

Both the compressibility and transient current measurements demonstrate the existence of two incompressible regions $(I)$ and $(II)$ 
corresponding to the pinning of the density in the central and guard regions respectively. It is interesting to know whether this pinning 
can be achieved if the microwave excitation is confined to the central and guard regions only. It is not possible to control the distribution 
of the microwave power inside the cavity directly, however we achieved some spatial selectivity by detuning the electrons from 
the intersubband resonance in the central and guard regions of the electron cloud. 
The results of compressibility experiments under these irradiation conditions are shown in Fig.~\ref{FigChiLoc}.

When the intersubband resonance conditions are realized only at the center of the cloud, 
an anomaly appears in the compressibility when the density in the center is around $n_{eD} \simeq 2 \times 10^6\;{\rm cm^2}$.
However if the intersubband resonance is confined to the guard region, 
the compressibility changes under irradiation only when the mean density in the guard
is around $n_{gD} \simeq  3 \times 10^6\;{\rm cm^2}$ regardless of the value of $n_{eD}$.
These observations confirm that the incompressible regions $(I)$ and $(II)$ appear predominantly owing to 
the excitation of the central and guard regions. Surprisingly, they also show that creating resonant conditions 
in those regions alone is not sufficient to create an incompressible state since the variations 
of the compressibility on Fig.~\ref{FigChiLoc} (typical $|\delta \chi|/\chi_0 \lesssim 0.1$) 
are much smaller than those in the case where the entire cloud was excited. Therefore, the formation of 
an incompressible state requires the excitation of the intersubband resonance in most of the 
electron cloud, confirming the existence of long-range correlations (on the ${\rm cm}$ scale) in this 
new state of the electron system.

The experiments performed with detuning in the guard show a large parameter range where $\delta \chi/\chi_0$ is only a function of the density in the center $n_{eD}$. This property is confirmed in Fig.~\ref{FigDosLoc}.a which compares the data displayed on Fig.~\ref{FigChiLoc} for different $N_e$. The absence of the strong hysteresis effect characteristic of the transition to the incompressible state allow to define two characteristic densities $n_{max}$ and $n_{min}$ (note that we checked that as expected when hysteresis is absent, the photocurrent and compressibility measurements coincide on the entire density and gate voltage range for this measurement) . The density $n_{max}$ corresponds to a maximum in $\delta \chi/\chi_0$ which is rather narrow with a half width quality factor of the maximum peak of around 10. We think that these characteristic densities correspond to the position of a density-dependent resonance in the system, which is probably a precursor to the formation of the incompressible state. Fig.~\ref{FigChiLoc}.b shows the dependence of $n_{max}$ and $n_{min}$ on the magnetic field, the observed quasi-linear dependence is consistent with the scaling proposed in the main text.

\clearpage 

{\bf Temperature dependence} 

\begin{figure}[h]
  \centerline{\includegraphics[clip=true,width=12cm]{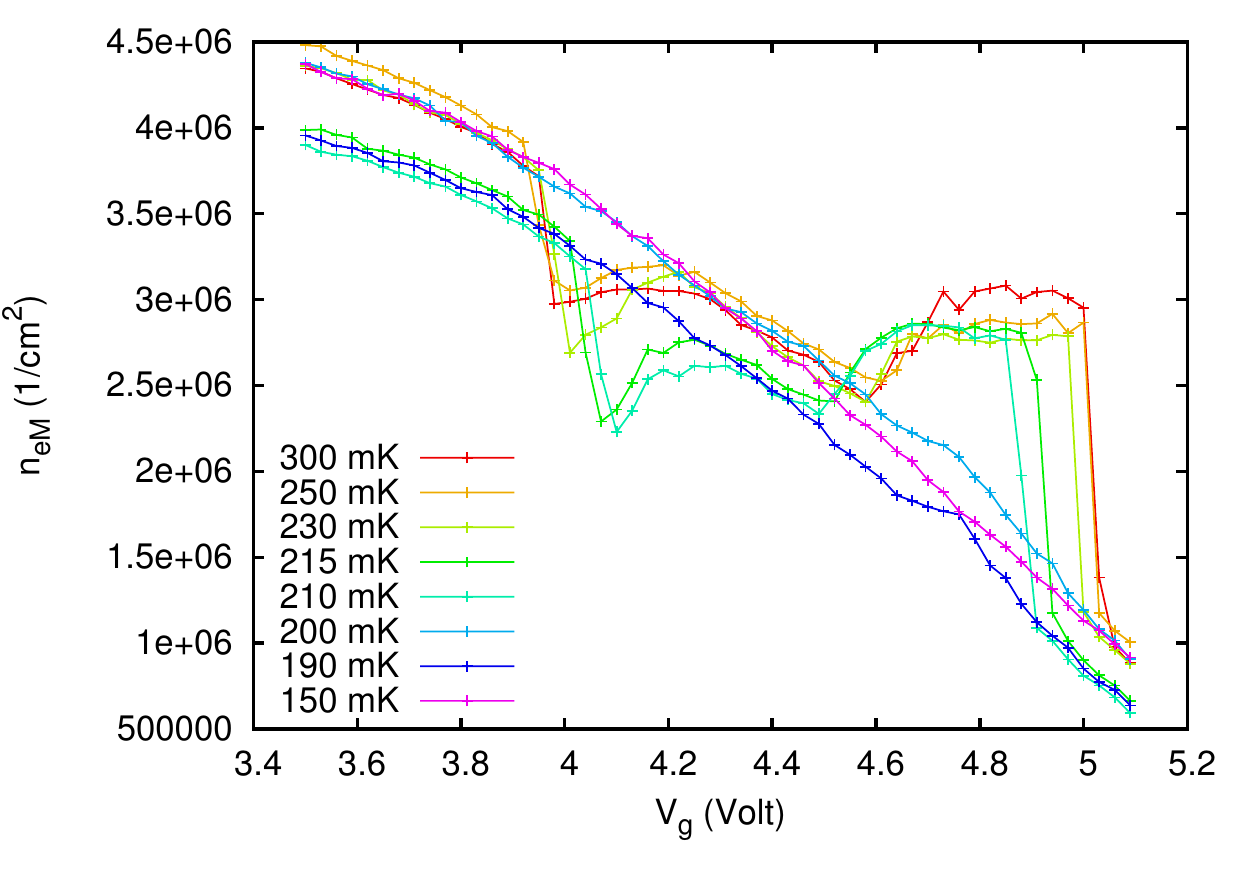}}
\caption{Density under irradiation as function of temperature combining data obtained for two different $N_e$ under the conditions of Fig.~5 (main text), the incompressible plateaux disappear at $T \le 200\;{\rm mK}$.
}
\label{FigTemp}
\end{figure}

We have attempted to experimentally study  the temperature dependence of the incompressible states. The results of these experiments are shown on Fig.~\ref{FigTemp}. One of the problems encountered in temperature dependent experiments was that the intersubband resonance shifts at lower temperatures effectively tuning the system out of resonance 
and destroying the incompressible states. We think that this frequency shift may be due to segregation of helium 3 impurity atoms onto the surface of helium 4 at low temperatures but further investigations are currently needed. Thus investigations of the temperature dependence require a careful calibration of the position of the intersubband resonance, highlighting the importance of the intersubband resonance to observe the formation of incompressible states.

\clearpage 

{\bf Extension of the photocurrent technique to the metastable incompressible branch}

We have shown (see Fig.~\ref{FigChi07}) that the incompressible state can be meta-stable and survive under continuous irradiation for parameters at which the electron-gas phase is stable under irradiation. In particular this occurs in region $(I)$ for increasing $V_g$ voltages. Since the compressibility also vanishes when the central region of the cloud is completely depleted, we do not know if the incompressible state extends as a metastable state beyond the depletion point or if it is destroyed close to the depletion voltage ($V_g \simeq 5.5\;{\rm Volt}$ for Fig.~\ref{FigChi07}). 

\begin{figure}[h]
\centerline{\includegraphics[clip=true,width=11.5cm]{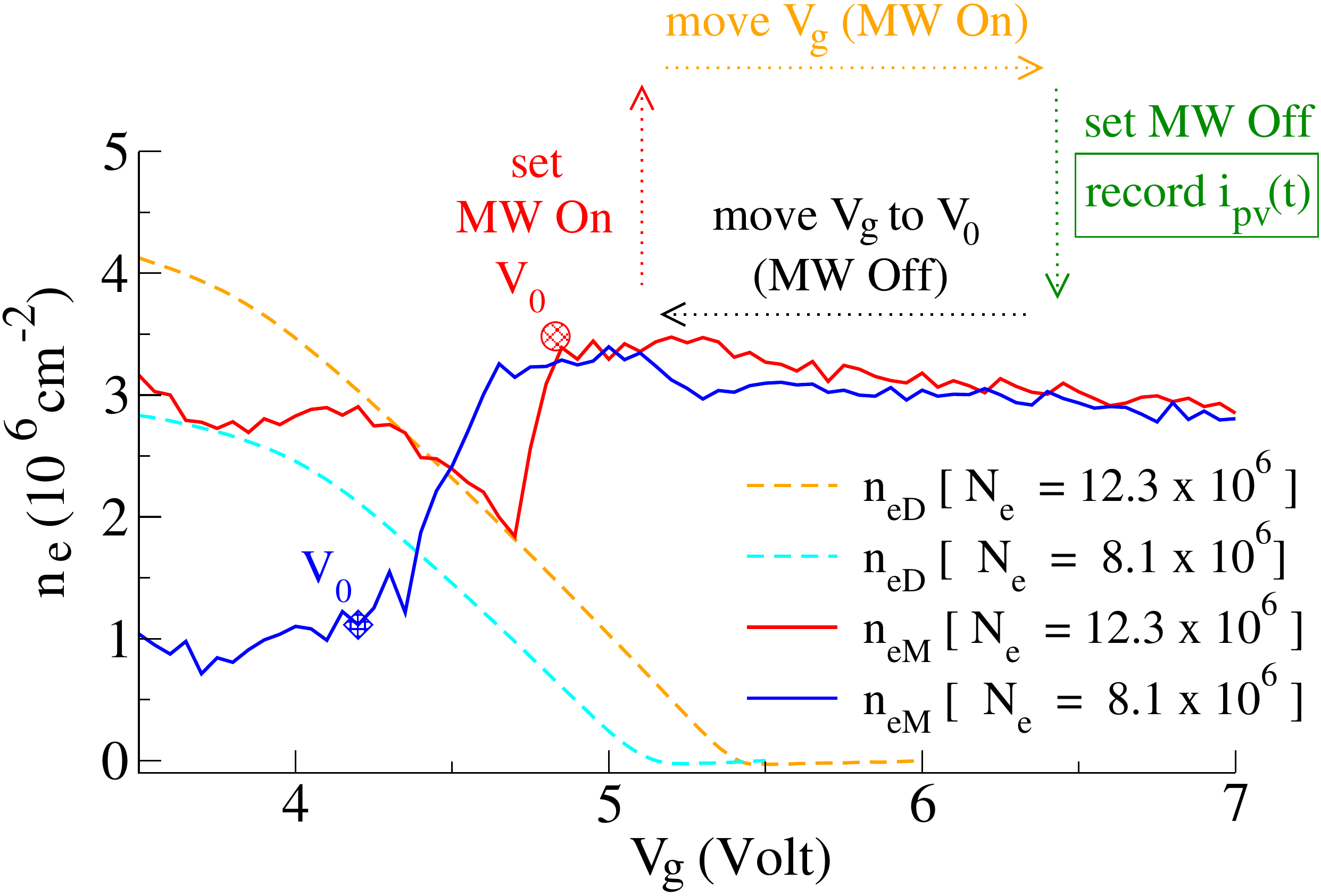}}
\caption{Modification of the photocurrent measurement to measure the electron density under irradiation even in the metastable regimes. For each guard voltage potential a cycle containing the following steps is performed: the cycle starts in the dark by setting $V_g = V_0$ where $V_0$ (indicated on the figure) is a voltage at which the incompressible state is the stable steady state of the system under irradiation. We turn on the microwaves, creating an incompressible state. Keeping the microwave irradiation, we move the guard voltage to its final value $V_g$ which is shown on the x-axis. The rate of change of the voltage is sufficiently slow  $0.1\;{\rm Volt/s}$ to maintain the system on the incompressible meta-stable branch. After the guard voltage has been set we wait for another $2\;{\rm s}$ and turn off the microwaves simultaneously triggering a photocurrent acquisition.  The relaxation of the surface electrons to their equilibrium density profile generates a current pulse $i_{pv}(t)$ which allows us to determine $n_{eM}$ using Eq.~(\ref{eq:dnepv}). At the end of the cycle the voltage is returned to its initial value $V_0$. The different steps of the cycle are summarized in the inset. 
The results of this experiment are shown for two values of $N_e$, the value $V_0$ is indicated by the symbols. 
Dashed lines show the equilibrium density $n_{eD}$ while the solid lines give the density under irradiation. Measurements were performed under the conditions of Fig.~5 in the main manuscript.
}
\label{FigCycle}
\end{figure}

\begin{figure}
\centerline{\includegraphics[clip=true,width=11.5cm]{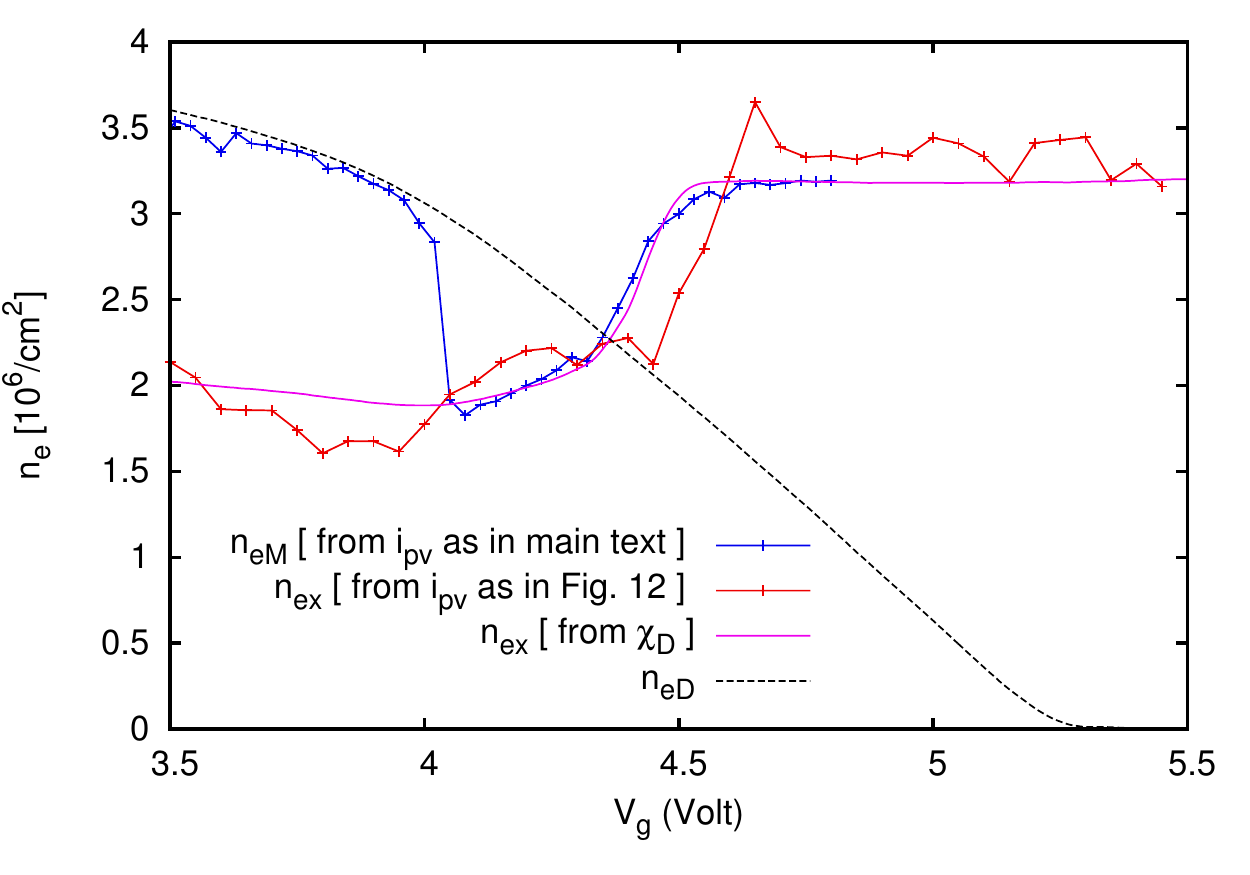}}
\caption{Density under irradiation obtained by three different techniques: photocurrent measurements, the modified photocurrent measurements described in Fig.~\ref{FigCycle} and integration of the compressibility under irradiation (in this case, the integration offset is chosen to match the values obtained from the other techniques in the high $V_g$ limit). Data is shown for $N_e = 10 \times 10^6$ and $J = 6.25$ under the conditions of Fig.~5 in the main manuscript.
}
\label{FigCycleCmp}
\end{figure}

To settle this issue we have modified the photocurrent technique to track the electronic density along the metastable incompressible branch under irradiation. This modification is described in Fig.~\ref{FigCycle} and shows that the metastable branch surprisingly extends far into the depletion region where all the electrons are expected to accumulate in the guard region. This shows that electrons can stay trapped in the center under irradiation even in presence of a potential dip in the guard exceeding $1 {\rm eV}$. As a final consistency check, we plot in Fig.~\ref{FigCycleCmp} the density under irradiation obtained using the three different measurements: the photocurrent method (Fig.~5 from the main manuscript and Fig.~\ref{FigIpv5.25} in supplementary materials), the modified photocurrent method described in Fig.~\ref{FigCycle} and the integration of the compressibility under irradiation using Eq.~(\ref{eq:dnechi}) and experimental data from Fig.~4 in the main manuscript). The good agreement between the three different procedures demonstrates that we have successfully reconstructed the two possible branches of the $n_e(V_g)$ dependence: the branch corresponding to the equilibrium electron gas and the excited state branch displaying incompressible behaviour. 

\clearpage 

{\bf Conductivity measurements}

\begin{figure}[h]
\centerline{\includegraphics[clip=true,width=8cm]{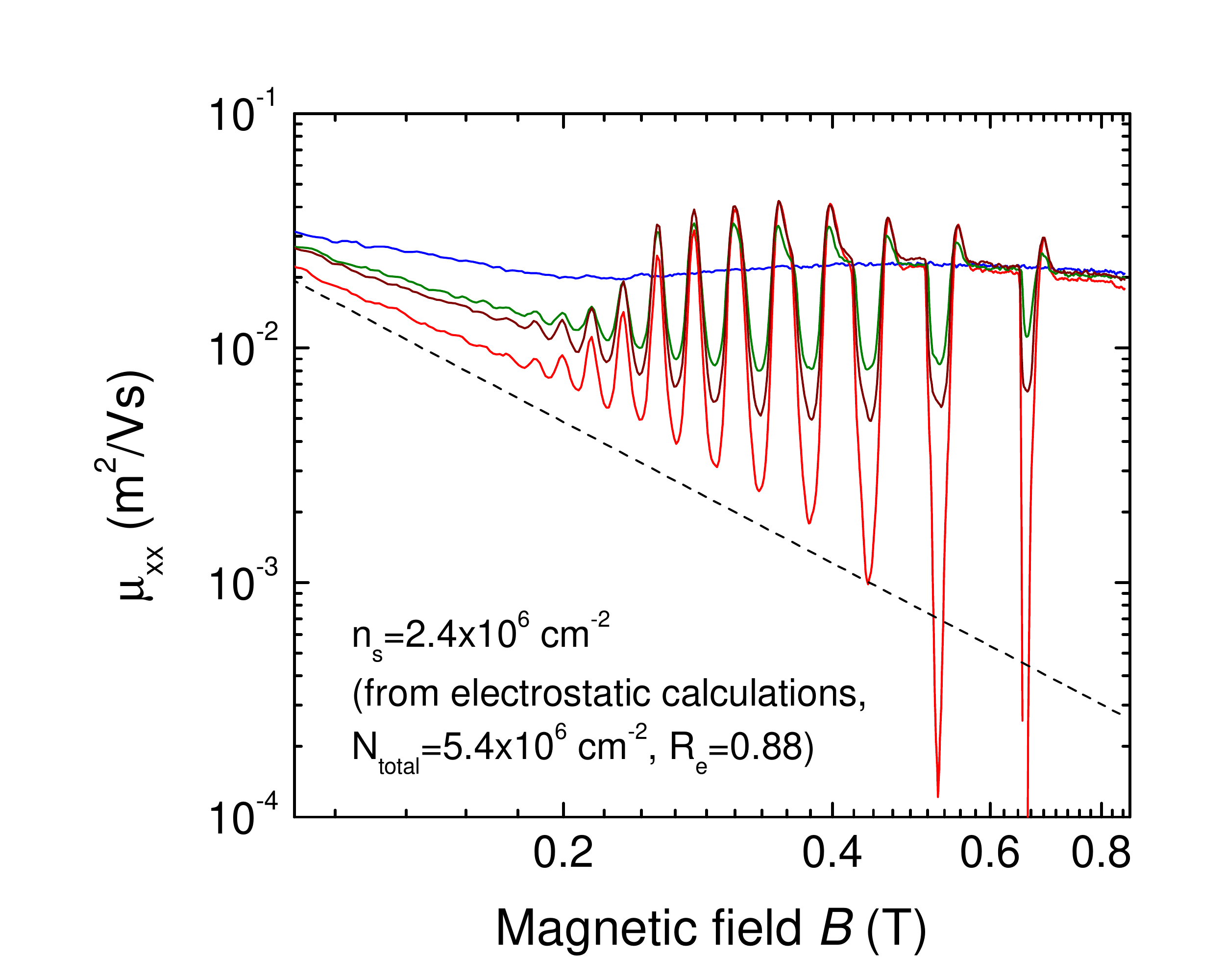}}
\caption{Mobility as function of magnetic field in the dark and under increasing microwave irradiation for a density of electrons $n_s = 2.5\times10^6{\rm cm^{-2}}$.
}
\label{FigMu1}
\end{figure}

\begin{figure}[h]
\centerline{\includegraphics[clip=true,width=8cm]{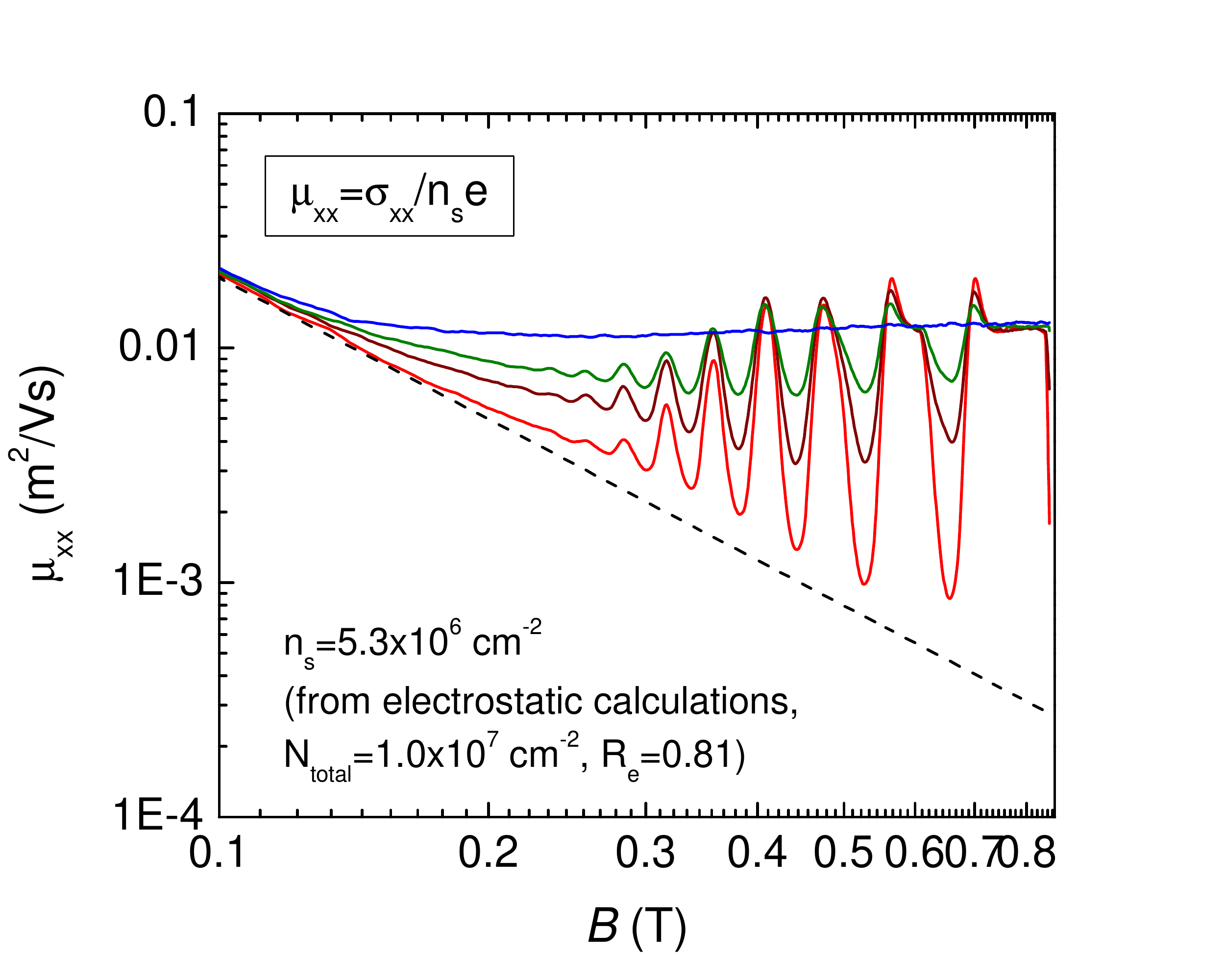}}
\caption{Mobility as function of magnetic field in the dark and under increasing microwave irradiation for a density of electrons $n_s = 5.3\times10^6{\rm cm^{-2}}$.
}
\label{FigMu2}
\end{figure}

Here we show some data on mobility under irradiation for different microwave powers. The dashed lines show the behavior expected from the Drude model, without irradiation the mobilities follow the Drude dependence at low magnetic fields but saturate to a fixed mobility at high magnetic field. This behavior corresponds to the quantum transport regime where $k_B T \ll \hbar \omega_c$ which was investigated in [ M. J. Lea, P. Fozooni, A. Kristensen, P. J. Richardson, K. Djerfi, M. I. Dykman, C. Fang-Yen, and A. Blackburn, {\it  Magnetoconductivity of two-dimensional electrons on liquid helium:Experiments in the fluid phase }, Phys. Rev. B {\bf 55}, 16280 (1997) ]. Microwave irradiation reduces the value of $\mu_{xx}$ for certain values of the parameter $\omega/\omega_c$ but the mobility remains above the Drude line even in this case for most traces. At the highest microwave power shown on Fig.~\ref{FigMu1}, the mobility abruptly drops to a very low value once it becomes close to $\mu_{xx} \simeq 10^{-3}\;{\rm m^2 V^{-1} s^{-1}}$, on Fig.~\ref{FigMu2} the mobility stays above this limit and the formation of ZRS is not observed.

This suggests that two independent mechanisms may combine to explain the formation of ZRS / incompressible phases: a mechanism that restores the conductivity to a value close to the Drude line at certain values of $J = \omega/\omega_c$ and an instability which occurs once $\mu_{xx}$ becomes sufficiently low and that leads to the formation of ZRS and incompressible behaviour. 

\end{document}